\documentclass[11pt,a4paper]{article}

\pdfoutput=1
\usepackage{jcappub}
\usepackage{times}

\usepackage{graphicx}
\usepackage{dcolumn}
\usepackage{bm}
\usepackage{color}
\usepackage{calrsfs}
\usepackage{hyperref}
\usepackage{multirow} 
\usepackage{float}
\usepackage{subfloat}
\input epsf

\newcommand{\be}{\begin{equation}}
\newcommand{\ee}{\end{equation}}
\newcommand{\bear}{\begin{eqnarray}}
\newcommand{\ear}{\end{eqnarray}}

\newcommand{\f}{\frac}

\begin{document}

\title{Warm dark matter signatures on the 21cm power spectrum: Intensity mapping forecasts for SKA}

\author[a]{Isabella P. Carucci,} 
\author[b,c]{Francisco Villaescusa-Navarro,}
\author[b,c]{Matteo Viel,}
\author[d,a,c,b]{Andrea Lapi}

\affiliation[a]{SISSA- International School for Advanced Studies, Via Bonomea 265, 34136 Trieste, Italy}
\affiliation[b]{INAF - Osservatorio Astronomico di Trieste, Via Tiepolo 11, 34143, Trieste, Italy}
\affiliation[c]{INFN sez. Trieste, Via Valerio 2, 34127 Trieste, Italy}
\affiliation[d]{Dip. Fisica, Univ. ÔTor VergataÕ, Via Ricerca Scientifica 1, 00133 Roma, Italy}

\emailAdd{ipcarucci@sissa.it}
\emailAdd{villaescusa@oats.inaf.it}
\emailAdd{viel@oats.inaf.it}
\emailAdd{lapi@sissa.it}

\abstract{We investigate the impact that warm dark matter (WDM) has in terms of 21cm intensity mapping in the post-reionization Universe at $z=3-5$.  We perform hydrodynamic simulations for 5 different models: cold dark matter and WDM with 1,2,3,4 keV (thermal relic) mass and assign the neutral hydrogen a-posteriori using two different methods that both reproduce observations in terms of column density distribution function of neutral hydrogen systems. Contrary to naive expectations, the suppression of power present in the linear and non-linear matter power spectra, results in an increase of power in terms of neutral hydrogen and 21cm power spectra. This is due to the fact that there is a lack of small mass halos in WDM models with respect to cold dark matter: in order to distribute a total amount of neutral hydrogen within the two cosmological models,  a larger quantity has to be placed in the most massive halos, that are more biased compared to the cold dark matter cosmology. We quantify this effect and address significance for the telescope SKA1-LOW, including a realistic noise modeling. The results indicate that we will be able to rule out a 4 keV WDM model with 5000 hours of observations at $z>3$, with a statistical significance of $>3\, \sigma$, while a smaller mass of 3 keV, comparable to present day constraints, can be ruled out at more than $2~\sigma$  confidence level with 1000 hours of observations at $z>5$.}


\maketitle

\section{Introduction}
\label{sec:introduction}

The $\Lambda{\rm CDM}$ cosmological model has been shown to be very successful in explaining a wide variety of observables: the anisotropies in the cosmic microwave background, the clustering of galaxies at low redshift, the abundance of galaxy clusters, the baryonic acoustic oscillations observed both in the spatial distribution of galaxies and in the Lyman-$\alpha$ forest. In this model, the dark matter is assumed to be cold (CDM), i.e. with negligible thermal velocities on all scales at high redshift. Although some tensions arise when combining different scales, a general picture emerge for which the simplest six parameter model provides a good fit to all these data sets \cite{Planck_2015}.

On small scales however, some predictions of the $\Lambda {\rm CDM}$ model seems to be at odds with observations. Among others, two major problems show up: the so-called \textit{core-cusp} problem and the subhalos abundance problem. The former arises because N-body simulations predicts the density profile of dark matter halos to be cuspy \cite{NFW, Aquarius}, whereas observations find them, in some galaxies, cored \cite{Salucci_2007, Gilmore_2007, Eymeren_2009, Naray_2010, Walker_2011}. The second problem comes because N-body simulations predicts, for Milky Way halos, the existence of a larger number of subhalos than the number of galaxy satellites found in the vicinity of our galaxy. 
More recently, this apparent crisis of the $\Lambda{\rm CDM}$ scenario has been casted in terms of the "too-big-to-fail" problem: namely the dynamical properties of the most massive subhalos of a Milky Way simulated halo are at odds with those of the observed satellites (dwarf galaxies), there must be a mechanism that prevent the star formation in this massive halos \cite{weinberg}. The most natural explanation would probably be to invoke some yet to be understood baryonic physical processes that deeply affect the properties at small scales \cite{ElZant_2001, Tonini_2006, Maccio_2012b}.

However, there is also a possible way to alleviate at least some small scales problems, while preserving the success of CDM on large scales, by assuming that the dark matter particles have intrinsic thermal velocities (although there are works that disagree, see for example \cite{Schneider_2014, Maccio_2013}). On one hand, the presence of thermal velocities will avoid the cuspy clustering of dark matter particles within gravitational potential wells, allowing the formation of cores \cite{Villaescusa-Navarro_Dalal, Maccio_2012, Shao_2013, Lovell_2014, Destri_2014} in the density profile, although most of the works have shown that an unrealistic low value of the WDM mass is needed to match observations. 
On the other hand, the collapse of dark matter halos can only take place for perturbations larger than the typical free-streaming length of the dark matter particles, thus, the presence of thermal velocities in the dark matter particles will naturally suppress the abundance of low mass subhalos \cite{Bode_2001,Avila-Reese_2001,Schneider_2012}.

While structure formation in a universe dominated by dark matter particles with large thermal velocities (hot dark matter) will proceed from top to bottom, the bottom-top paradigm is preserved in the case of dark matter particles with relatively low thermal velocities that could be considered as warm dark matter (WDM). WDM candidates comprise sterile neutrinos and gravitinos or other thermal relics.

Given the fact that WDM can not substantially cluster on scales smaller than its free streaming scale, the matter power spectrum in a cosmological model with WDM will exhibit, on small scales, a cut-off. Because the Lyman-$\alpha$ forest can be used to probe the shape and amplitude of the matter power spectrum on small scales, the current tightest constrains on the WDM mass come from such observations: $m_{\rm WDM}\geqslant3.3~{\rm keV}(2\sigma)$ \cite{Viel_2013}. 

Nowadays, our constraints on the matter power spectrum arise from different probes such as the galaxy clustering, weak lensing, abundance of galaxy clusters and the Lyman-$\alpha$ forest. In the near future, observations of the redshifted 21cm radiation from neutral hydrogen will allow us to constrain the matter power spectrum on redshifts not accessible with the above observables with an unprecedented precision \cite{Bull_2014, Camera_2013}.  The idea is to study the statistical properties in the integrated emission from unresolved galaxies using low angular resolution surveys; this technique is called intensity mapping.
This observable is promising: it is at high redshift and thereby should be closer to sample the linear cut-off in the matter power spectrum induced by WDM; the scales and redshifts addressed here should only be marginally affected by baryonic processes like feedback; the total amount of neutral hydrogen is constrained by data, i.e. absorption lines. 

In order to extract the maximum information from intensity mapping surveys, it is critical to model the spatial distribution of neutral hydrogen. In this paper we investigate the differences in the spatial distribution of neutral hydrogen between models with cold and warm dark matter. Since the quantity directly probed by radio-telescopes is the 21cm power spectrum, we study the signatures imprinted by WDM on this observable. We do so by running high resolution hydrodynamic simulations. We also compute the sensitivity with which the future Square Kilometre array\footnote{https://www.skatelescope.org/} (SKA) will constrain the 21cm power spectrum, and therefore, how much it will be able to discriminate among different cosmological models.

The scientific motivation behind this work is actually more general than the WDM framework we are investigating since we are aiming at providing a quantitative interpretation of the clustering signal in intensity mapping in a completely new regime of scales and redshifts, with very little (if any) overlap
with other observational probes.

We notice that the signatures left by WDM on the 21cm power spectrum has already been studied in \cite{sitwell}, which focused their study in the reionization era, with its richer astrophysical processes while in this work we investigate the post-reionization era. Furthermore, the impact of WDM on the higher redshift regime has been recently explored by \cite{dayal,maio15}, and more exotic annihilating DM and decaying WDM are taken into consideration by \cite{Evoli_2014}.

This paper is organized as follows. In Sec. \ref{sec:simulations} we present the hydrodynamic simulations  used in the present work. In this section we also study the differences in the spatial distribution of matter between the models with CDM and WDM. The two different methods used to simulate the spatial distribution of neutral hydrogen are described in Sec. \ref{sec:HI_distribution}, while the spatial distribution of neutral hydrogen in the models with CDM and WDM is investigated in Sec. \ref{sec:results}. The capability of the future SKA1-LOW array to distinguish between the different models is fully quantified in Sec. \ref{sec:results}. Finally, a summary of the work carried out in this paper and the main conclusions of it are presented in Sec. \ref{sec:conclusions}.

\section{Simulations and matter distribution}
\label{sec:simulations}

In this section, we present first the hydrodynamic simulations in subsection \ref{subsec:simulations}, whereas the impact of WDM on the spatial distribution of matter, studied in terms of the matter power spectrum and the halo mass function is discussed in subsection \ref{subsec:matter}.

\subsection{Hydrodynamic simulations}
\label{subsec:simulations}

Our simulation suite comprises a set of 5 high-resolution hydrodynamic N-body simulations run using the TreePM+SPH code {\sc GADGET-III} \cite{Springel_2005}. We follow the evolution of $512^3$ CDM/WDM and $512^3$ baryon (gas+stars) particles within simulation boxes of comoving sizes equal to 30 $h^{-1}{\rm Mpc}$. The mass resolution of our simulations is $m_{\rm DM}=1.50\times10^7~h^{-1}M_\odot$ and $m_{\rm b}=2.74\times10^{6}h^{-1}M_\odot$. The gravitational softening is set to $1/40$ of the mean inter-particle linear spacing, i.e. 1.5 $h^{-1}{\rm kpc}$. We have simulated five different cosmological models: 1 model with CDM and 4 models with WDM, each of them with a different Fermi-Dirac mass of the WDM particles: 1 keV, 2 keV, 3 keV and 4 keV (thermal relic mass, see below). The values of the cosmological parameters are the same in all the models and in agreement with the latest results of the Planck satellite \cite{Planck_2015}: $\Omega_{\rm m}=0.3175$, $\Omega_{\rm b}=0.049$, $\Omega_\Lambda=0.6825$, $h=0.6711$, $n_s=0.9624$ and $\sigma_8=0.834$.  We notice that the resolution on our simulations 
is high enough to provide converged results when the HI is assigned either with the halo or particle based method (see \cite{Villaescusa_2014a}).

Star formation is modeled using the effective multi-phase model of Springel \& Hernquist \cite{Springel-Hernquist_2003}. The code also simulates radiative cooling by hydrogen and helium and heating by a uniform Ultra Violet (UV) background. Both the cooling routine and the UV background have been modified to obtain a desired thermal history, corresponding to the reference model of \cite{viel13}, which is kept fixed in the following.

The initial conditions were generated at $z=99$ using the Zel'dovich approximation. For the model with CDM the matter power spectrum and the transfer functions were computed using CAMB \cite{CAMB} whereas for the models with WDM the power spectra, $\tilde{P}_a(k)$, were calculated as
 \cite{Viel_WDM_2012}
\be
\tilde{P}^{\rm \Lambda WDM}_a(k)=T^2_{\rm lin}(k)P^{\rm \Lambda CDM}_a(k)
\ee
where $a$ stands for either baryonic and the non-baryonic matter components and with 
\be
T_{\rm lin}(k)=\left(1+(\alpha k)^{2\nu}\right)^{-5/\nu}
\label{eq:T2}
\ee
where $\nu=1.12$ and 
\be
\alpha(m_{\rm WDM})=0.049\left(\frac{1~{\rm keV}}{m_{\rm WDM}}\right)^{1.11} \left(\frac{\Omega_{\rm WDM}}{0.25} \right)^{0.11} \left(\frac{h}{0.7}\right)^{1.22}~h^{-1}{\rm Mpc}~.
\ee
Furthermore, for the cosmological models with WDM we have added thermal velocities, on top of the peculiar velocities, to the non-baryonic particles. For a given WDM particle, the modulus of the thermal velocity vector is drawn randomly from a Fermi-Dirac distribution with a mean equal to \cite{Bode_2001}
\be
\bar{V}_{\rm WDM}(z)=0.012(1+z)\left(\frac{\Omega_{\rm WDM}}{0.3}\right)^{1/3}\left(\frac{h}{0.65}\right)^{2/3}\left(\frac{{\rm keV}}{m_{\rm WDM}}\right)^{4/3}~{\rm km/s}
\ee
and the direction of the velocity vector is taken randomly.  We analyzed snapshots at $z=3,3.5,4,4.5,5$ and identified the dark matter halos and subhalos using the Friends-of-Friends (FoF) \cite{FoF} and {\sc SUBFIND} \cite{Subfind,Dolag_2009} algorithms, respectively.

\begin{figure}
\includegraphics[width=\textwidth]{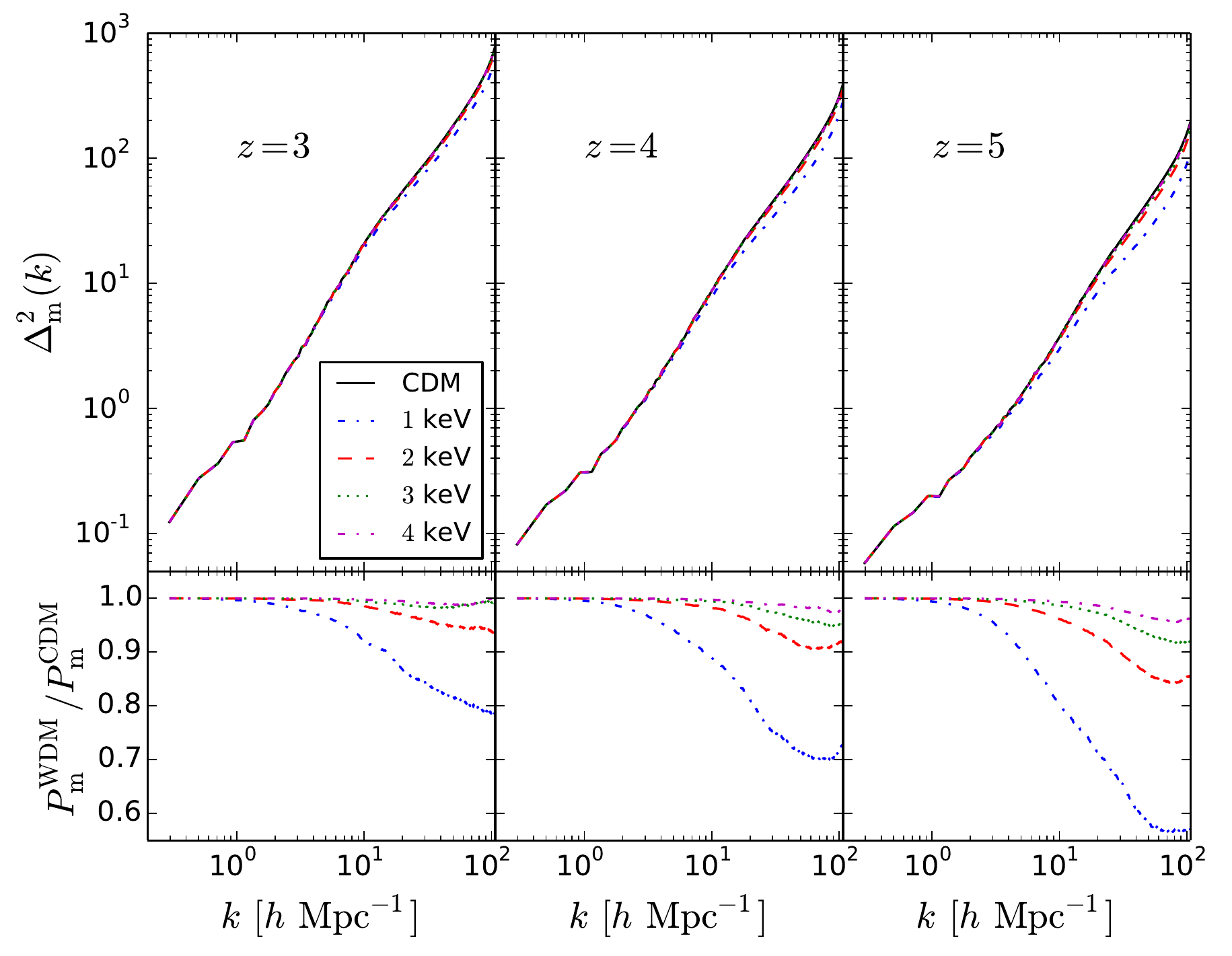}
\caption{Dimensionless power spectrum of total matter (DM + baryons), $\Delta^2_{\rm m}(k)=k^3 P_{\rm m}(k) / (2 \pi^2)$, at redshifts $z=3$ (left), $z=4$ (middle) and $z=5$ (right). In the top panels we plot with a solid black line the matter power spectrum of the model with CDM and with colored lines the results for the WDM models: $1$ keV (dash-dotted blue), $2$ keV (dashed red), $3$ keV (dotted green) and $4$ keV (dash-dotted magenta). The bottom panels show the relative differences between the WDM and the CDM models. We set the $k$ range up to  the Nyquist frequency $(k\simeq 107\, h\, {\rm Mpc}^{-1} )$.}
\label{fig:matterPk}
\end{figure}

\subsection{Impact of WDM on the matter distribution}
\label{subsec:matter}

We now examine the impact of WDM on the spatial distribution of matter. In particular, we focus our attention on the matter power spectrum and the halo mass function.
This high redshift regime $z=3-5$ is relatively new for probing WDM models in the sense that it partially overlaps with Lyman-$\alpha$ forest data, while is lower compared to 
the reionization and early structure formation regimes investigated by \cite{sitwell,dayal,maio15}.

In the top panels of figure \ref{fig:matterPk} we show the total matter power spectrum, $P_{\rm m}(k)$, for each of our five different cosmologies. The bottom panels represent instead the relative differences between the WDM and the CDM models. Since WDM can not substantially cluster on scales smaller than its free-streaming length, the matter power spectrum in WDM cosmologies presents a relatively sharp cut-off on small scales; as expected, this suppression, while being less pronounced that the linear cut-off, is larger at higher redshift and for smaller WDM masses. The clustering on scales larger than the free-streaming length is not affected by the thermal velocities of the WDM particles, and thus, the amplitude of the matter power spectrum is the same for all the models.

\begin{figure}
\includegraphics[width=\textwidth]{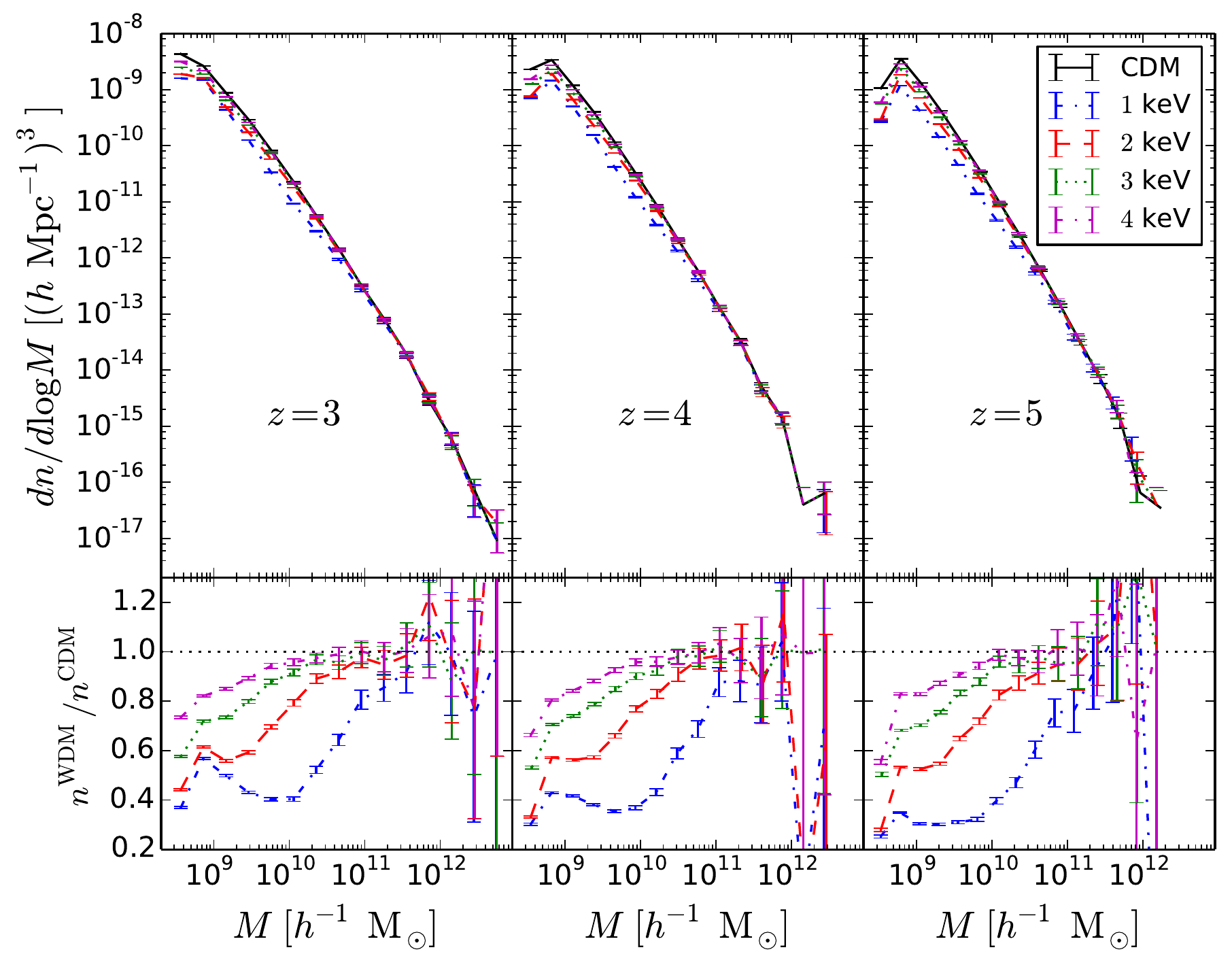}
\caption{Halo mass function (using the FoF catalogue) at redshifts $z=3$ (left), $z=4$ (middle) and $z=5$ (right). In the top panels the halo mass function for the CDM model is shown with a solid black line whereas the colored lines represent the results for the WDM models: $1$ keV mass (dash-dotted blue), $2$ keV (dashed red), $3$ keV (dotted green) and $4$ keV (dash-dotted magenta). The bottom panels show the relative differences between the WDM and the CDM models. Error bars represent the uncertainty in the halo mass function assuming that the number of halos follows a Poissonian distribution.}
\label{fig:MF}
\end{figure}

In figure \ref{fig:MF} we show the halo mass function for the different models investigated in this paper. In the top panels we plot the halo mass function at $z=3$ (left), $z=4$ (middle) and $z=5$ (right). The error bars represent the errors on the halo mass function assuming that the number density of halos follows a Poissonian distribution. In the bottom panel, we plot the relative difference between the WDM models and the model with CDM. As it can be seen in the figure, WDM induces a suppression on the abundance of low-mass dark matter halos, with respect to the abundance of halos in the CDM model, which increases with redshift and decreases with the mass of the WDM particles. As we shall see later, this effect will drive the changes we observe in the 21cm power spectrum between the models with CDM and WDM.

In deriving the mass function of WDM simulations, spurious fragmentation could produce an increase of halos at a low mass scale, see for example \cite{Schneider_2012}. Our simulations, except the model with 1 keV, are not affected by such effect, both because we artificially boost late time thermal velocities at the beginning of our simulations \cite{Schneider_2013} and because we do not reach enough mass resolution for our WDM models, which have higher masses compared to those in the literature. This can be directly seen from the behaviour of the curves in figure \ref{fig:MF}, that do not show any characteristics steep power law upturn for low halo masses. Moreover, as shown in \cite{Schneider_2013}, the mass function of the artificial clumps strongly depends on the resolution of the simulations; we checked the mass functions resulting from simulations of $1/8$ the resolution we have (i.e. comoving box size of 60 $h^{-1}{\rm Mpc}$) and we have verified the convergence for small halo masses.

Overall, for WDM masses between 3 and 4 keV we observe a suppression in power up to $\sim10\%$ on very small scales and a reduction in the number of $10^9~h^{-1}M_{\odot}$ halos of the order of 20-40\% compared to the CDM case. 

We fit the WDM mass functions obtained from the simulations (FoF catalogue) by parametrizing their deviation from the mass function predicted by the Sheth \& Tormen \cite{Sheth-Tormen} formula\footnote{This is achieved by computing the Sheth \& Tormen mass function using the linear power spectrum of the WDM model.} 
\be
\frac{dn_{\rm WDM}^{\rm sim}}{d\log M}(M)=\frac{dn^{\rm ST}_{\rm WDM}}{d\log M}(M) \left[1 - \alpha e^{- M / M_0}\right]\,,
\label{eq:MF}
\ee
where $\alpha$ and $M_0$ are free parameters. The best fit values for the different WDM models and redshifts are shown in the table \ref{tab:MF_fit}. We find that the mass function of the 1 keV WDM cosmology is the only one that is under predicted by the Sheth \& Tormen model. We exclude the possibility of fragmentation, because $n_{\rm WDM}^{\rm sim}$ lacks the associated steeper raise for low halo masses (as already discussed above), although this effect could point to a transitional and physical regime before fragmentation takes place (see figure B1 in \cite{Schneider_2013}). We also find that the mass function of the CDM model is over predicted by the Sheth \& Tormen formula, although only in the low mass end; the best fit values obtained using equation \ref{eq:MF} are shown in the table \ref{tab:MF_fit} with $\infty$ as the value of the WDM mass.

\begin{table}
\begin{center}
\begin{tabular}{| c | c | c | c |}
\hline
$m_{\rm WDM}\,[{\rm keV}]$ & $  z  $ & $\alpha$ & $M_0 \, [10^9\, h^{-1}{\rm M}_{\odot}]$\\  
\hline \hline
\multirow{3}{*}{1}  & 3  & -1.72	  & 0.887 \\ \cline{2-4}
 &  4  & -1.92	 &  0.960 \\ \cline{2-4}
 &  5  & -2.06	 &  0.737 \\ \cline{2-4}
\hline
\multirow{3}{*}{2}  & 3 &  0.571 & 	 3.98 \\ \cline{2-4}
 & 4  & 0.389	 &  3.80 \\ \cline{2-4}
  & 5 &  0.286 & 	 4.81 \\ \cline{2-4}
\hline
\multirow{3}{*}{3}  & 3  & 0.830	 &  1.86 \\ \cline{2-4}
  & 4  & 0.726   &   1.28 \\ \cline{2-4}
  & 5  & 0.427  &    1.74 \\ \cline{2-4}
\hline
\multirow{3}{*}{4}  & 3  & 0.958   &   1.44 \\ \cline{2-4}
 &  4  & 1.04    &   0.890 \\ \cline{2-4}
 & 5  & 0.522  &    1.05 \\ \cline{2-4}
\hline
\multirow{3}{*}{$\infty$}  & 3  & 1.26	  & 1.04 \\ \cline{2-4}
 &  4  & 2.77	 &  0.491 \\ \cline{2-4}
 &  5  & 1.69	 &  0.455 \\ \cline{2-4}
 \hline
\end{tabular}
\caption{\label{tab:MF_fit}Parameter values for the mass function fit of equation \ref{eq:MF}.}
\end{center}
\end{table}

\section{Modeling the neutral hydrogen distribution}
\label{sec:HI_distribution}

In this section we describe the two different approaches we use to model the spatial distribution of neutral hydrogen. In subsection \ref{sub:bagla} we present the {\it halo based} method, which relies on two assumptions: no HI is found outside dark matter halos and the amount of neutral hydrogen inside the halos is a function of their total mass only. The so-called {\it particle based} method is summarized in subsection \ref{sub:dave}: according to this modeling, HI is assigned to every single gas particle in the simulation, depending on the physical properties of the particles themselves.
Both models have been extensively investigated for intensity mapping at similar redshifts in \cite{Villaescusa_2014a} and are meant to bracket conservatively the possible ways of populating halos with HI.

\subsection{Halo based method}
\label{sub:bagla}

\begin{figure}
\centering
\includegraphics[width=0.7\textwidth]{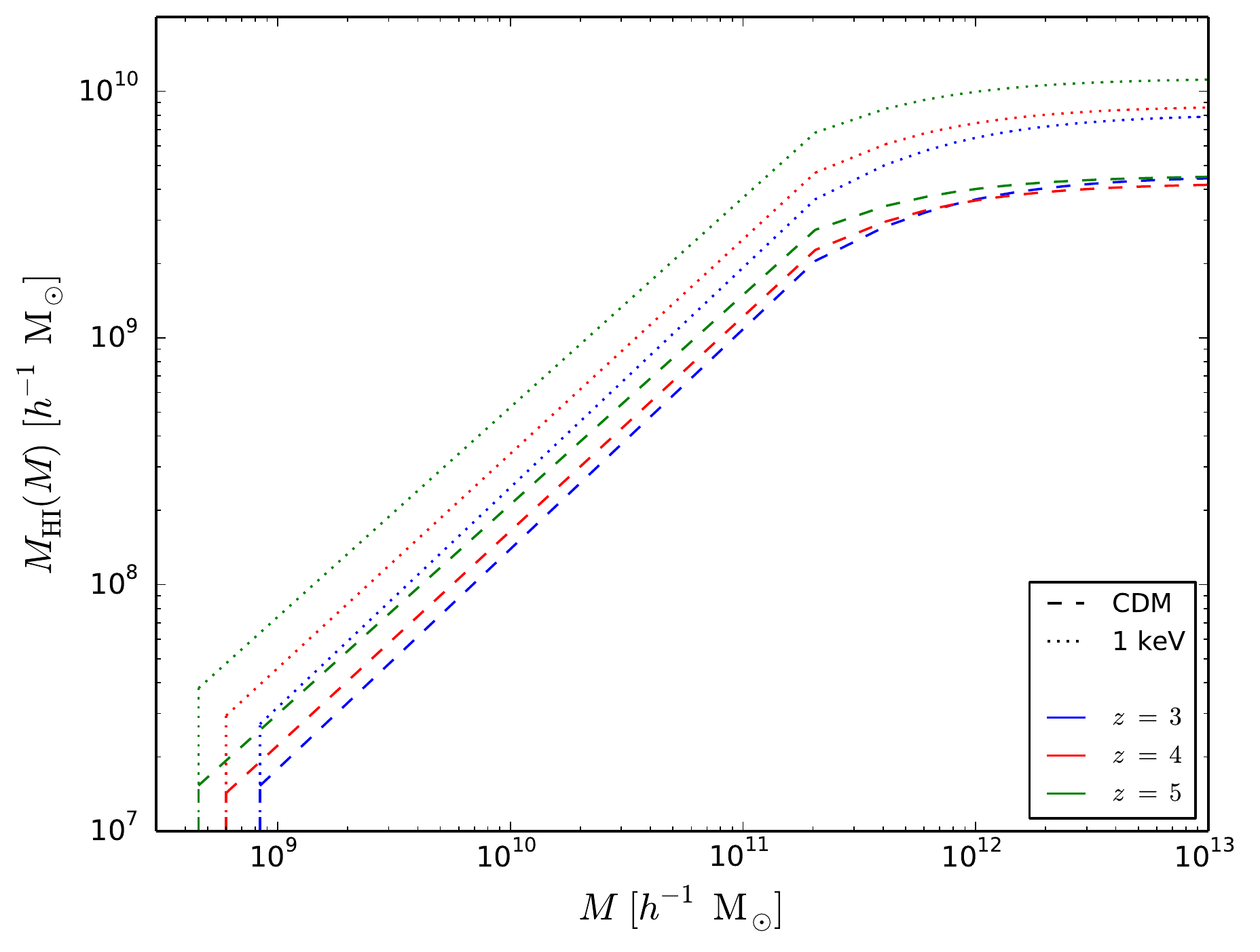}
\caption{The function $M_{\rm HI}(M)$ relating the HI mass and total mass of a dark matter halo of mass $M$, employed in the halo based model as described by the equation \ref{eq:bagla}, for the CDM (dashed lines) and 1 keV WDM (dotted lines) cases, for redshifts $z=3, 4, 5$ (blue, red and green). The values of the parameter $f_3$ are 0.0183, 0.0239,  0.0337 for the CDM, redshift 3 to 5, and 0.0325, 0.0491, 0.0835 for the 1 keV, redshift 3 to 5.}
\label{fig:bagla}
\end{figure}

Here we follow one of the models described in Bagla et al. \cite{Bagla_2010}. Starting from the snapshots of our N-body simulations, we first single out the dark matter halos and all the particles belonging to them, as identified by the Friends-of-Friends algorithm (FoF catalogue). We  then assume that the amount of HI hosted by a dark matter halo of mass $M$ is given by the function $M_{\rm HI}(M)$, i.e. we neglect any environmental effects. We then split the total HI mass of the dark matter halo equally among all its gas particles. Following the model 3 of Bagla et al., we model the function $M_{\rm HI}(M)$ as:
\be
  M_{\rm HI}(M)=\begin{cases}
    f_3 \frac{M}{1+(M/M_{\rm max})} & \text{if $M_{\rm min}\leq M$}\\
    0 & \text{otherwise}.
  \end{cases}
 \label{eq:bagla}
\ee
The values of the free parameters $M_{\rm max}$ and $M_{\rm min}$ are chosen in correspondence dark matter halo circular velocities $v_{\rm circ}$ of $200$ and $30$ km ${\rm s}^{-1}$ respectively, using the relation
\be
M = 10^{10} {\rm M}_{\odot} \left( \frac{v_{\rm circ}}{60 \,{\rm km\,s}^{-1}} \right)^3 \left( \frac{1+z}{4} \right)^{-1.5} ~,
\ee
derived from virialisation arguments. The lower mass cut-off $M_{\rm min}$ takes into account that a minimum hydrogen density (clustered in a minumum halo potential well) is needed to have self-shielding and prevent the gas to be fully ionised. The fall-off above $M_{\rm max}$ is introduced to mimic observations: galaxies in dense environments as galaxy clusters are HI-poor. 
For further details on this model we refer the reader to \cite{Bagla_2010}.
The magnitude of $M_{\rm min}$ is of the order of $10^9$ $h^{-1} {\rm M}_{\odot}$, so to distribute HI consistently, our simulations need to resolve dark matter halos up to this mass halos or lower. We achieve this by setting the box size of our simulations to 30 $h^{-1}{\rm Mpc}$, among the highest values possible considering the number of particles, $2\times512^3$, we are dealing with.

We are left with a single free parameter, $f_3$, whose value is chosen by demanding that the total amount of HI in our boxes reproduces the observational constraints arising from the abundance of DLAs at high redshift: $\Omega_{\rm HI}(z) = 10^{-3}$, i.e. we impose:
\be
f_3 \sum_{i=0}^n \frac{M_i}{1+\frac{M_i}{M_{\rm max}}}\Theta(M_i-M_{\rm min}) = \Omega_{\rm HI} L^3 \rho_{\rm c}^0 \,,
\ee
where $L=30~h^{-1}{\rm Mpc}$ is the simulation box size, $ \rho_{\rm c}^0 = 2.78 \times 10^{11} h^2 {\rm M_{\odot} Mpc^{-3}}$ is the present day critical density of the universe, $\Theta(x)$ is the Heaviside step function and the index $i$ run over all the dark matter halos of the simulation. We stress that by defining $f_3$ in this way for each snapshot and simulation, we assure to have the same reference value of $\Omega_{\rm HI}$ independently of the redshift and cosmology we analyze, within the halo based method. We will show that this is not the case in the particle based method. 

We show the function $M_{\rm HI}(M)$ that the halo based method prescribes for the models with CDM and $1$ keV WDM in figure \ref{fig:bagla}. The $M_{\rm HI}(M)$ function steeply increases up to $M_{\rm max}$, then displays a plateau. The values of $M_{\rm min}$ and $M_{\rm max}$ do not change among different cosmologies, but the normalization (through the parameter $f_3$) does: it gets higher for warmer DM scenarios.

\subsection{Particle based method}
\label{sub:dave}

By employing this method we assign neutral hydrogen to all gas particles in the simulation according to their physical properties. Compared to the halo based method, we do not rely on any definition of dark matter halo or on any assumption on the amount of HI outside halos and therefore we can also predict the amount of HI in places different to halos as filaments and cosmic voids.

We proceed as depicted in \cite{Dave_2013}: for every single gas particle in the simulation we compute the neutral hydrogen fraction in photo-ionization equilibrium with the Ultra Violet (UV) background and correct that fraction to account for both HI self-shielding and formation of molecular hydrogen. Hydrogen of star forming particles is split in totally neutral for the cold phase, and totally ionized for the hot phase, i.e. we consider than the HI/H fraction of star forming particles in photo-ionization equilibrium is equal to its multi-phase cold gas fraction. The strength of the UV background, at a particular redshift, is corrected by demanding that the mean flux of the Lyman-$\alpha$ forest reproduces the observations \cite{Becker_2013}, with the approach described in \cite{Villaescusa_2014a}. For every gas particle we then compute the radial column density profile $N_{\rm HI}(r)$ making use of the SPH spline kernel $W(R)$:
\be
N_{\rm HI}(r)=\frac{0.76 \,m}{m_{\rm H}} \left(\frac{\rm HI}{\rm H}\right) \int_{r}^{r_{\rm SPH}} W(r') dr' \,,
\ee
with $m/m_{\rm H}$ being the gas particle mass in units of hydrogen atom mass, HI/H is the neutral hydrogen fraction in photo-ionization equilibrium and $r_{\rm SPH}$ the SPH smoothing length. If exists a radius $r$ such that $N_{\rm HI}(r) = 10^{17.2}$ cm$^{-2}$, then the sphere from $R=0$ to $R=r$ is considered to be self-shielded agains the external UV radiation and its HI/H fraction is set to $0.9$. For the surrounding spherical shell $(R>r)$ it still holds the neutral fraction HI/H coming from the photo-ionization equilibrium assumption. 

The last step consists in correcting the HI/H fraction computed in the previous step to account for the formation of molecular hydrogen. Using the density and internal energy, we compute the pressure of gas particles $P$ and use it to correct the HI/H fraction computed previously using the observed relation between the surface densities of HI and H$_2$ with the pressure of the disk galaxy \cite{Blitz_2006}:
\be
R_{\rm mol} = \frac{\Sigma_{\rm H_2}}{\Sigma_{\rm HI}}= \left( \frac{P/k_{\rm B}}{3.5 \times 10^4 \,{\rm cm}^{-3}{\rm K}} \right)^{0.92}
\label{eq:H2}
\ee
with $k_{\rm B}$ the Boltzmann constant. H$_2$ is assigned only to star forming particles. We note that we are implicitly assuming that the above relationship, which arises from observations at low-redshift, holds also at high redshift which may not be the case (see for instance \cite{Dave_2013, Krumholz_2011}).

\section{Results}
\label{sec:results}

\begin{figure}
\includegraphics[width=\textwidth]{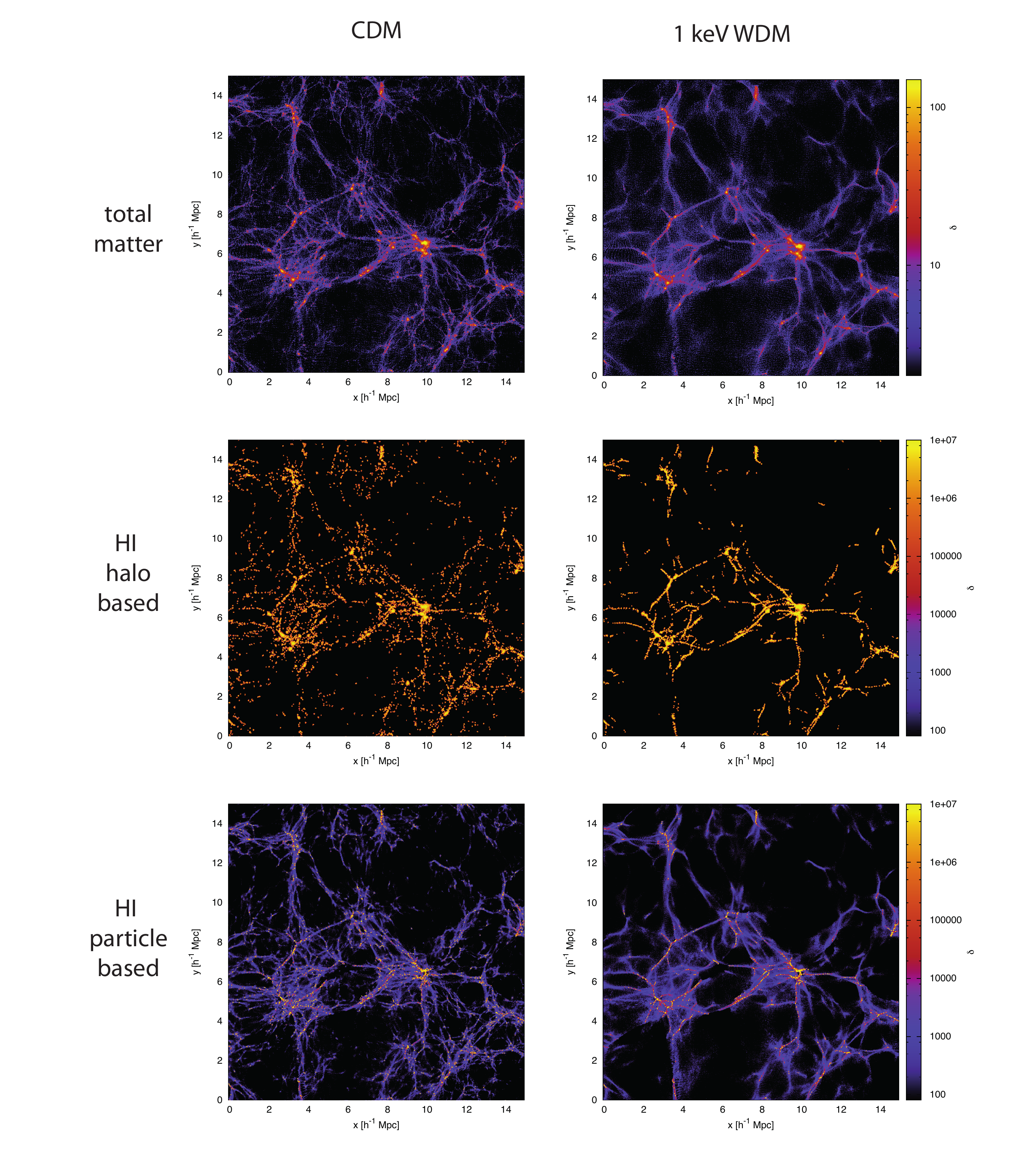}
\caption{Spatial distribution, at redshift $z = 3$, of the density contrast of total matter (DM + baryons) and of the HI placed according to the halo and particle based model, upper, middle and low row respectively, for the CDM (left column) and 1 keV WDM (right column)  scenarios. We zoom in a $(15\, h^{-1}  {\rm Mpc})^2$ region, taking a slice of $2\, h^{-1}  {\rm Mpc}$ width.}
\label{fig:images}
\end{figure}

In this section we use the methods described above to model the spatial distribution of neutral hydrogen in our five different cosmological models.

An overall picture of the spatial distribution of matter and HI analyzed in this work is given in figure \ref{fig:images}, where we confront the CDM scenario (left column) with the 1 keV WDM (right column) at redshift $z=3$. The difference among the two cosmologies can be noted in the total matter distribution (top panels), where the CDM displays much more clustered structures while the spatial distribution of HI also differ in the two cosmologies (middle and bottom panels). By using the halo based, HI is present just in halos (middle panels), while by employing the particle based (bottom panels), the HI distribution is smoother since HI is assigned to every single gas particle in the simulation. The differences in the neutral hydrogen clustering properties can also be noticed by visual inspection.

We perform a more quantitative analysis of the HI distribution. First, in the subsection \ref{sub:fHI}, we compare the HI column density distribution function that we obtain, for each cosmological model, with observations. In the subsections \ref{sub:Pk_HI} and \ref{sub:bias_HI} we investigate the differences in terms of HI power spectrum and of the HI bias $b^2_{\rm HI}(k)=P_{\rm HI}(k)/P_{\rm m}(k)$. The 21cm power spectrum is computed in the subsection \ref{sub:21Pk}, where we also investigate the accuracy with which the SKA1-LOW telescope will be able to measure it.

\subsection{HI column density distribution function}
\label{sub:fHI}

\begin{figure}
\includegraphics[width=\textwidth]{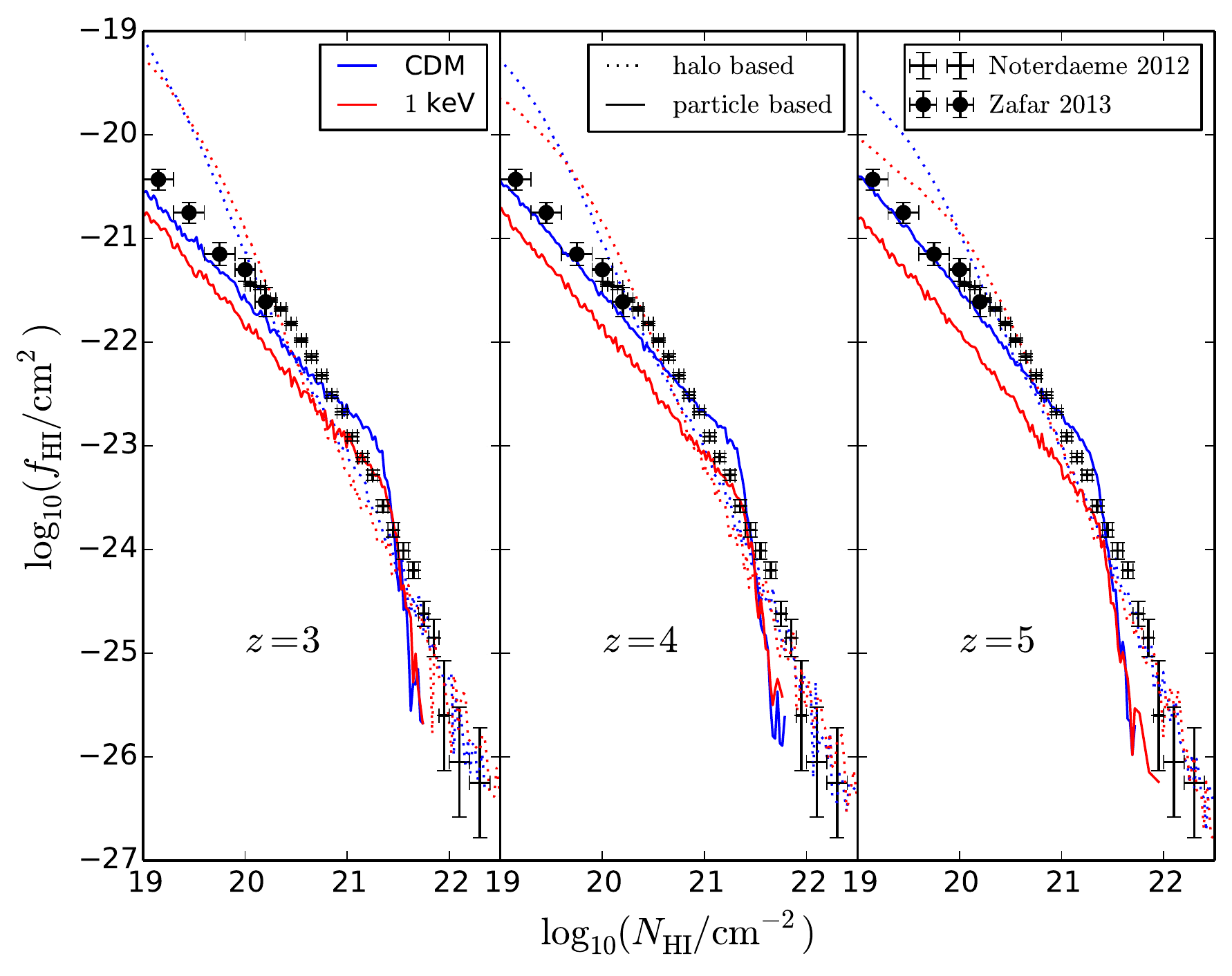}
\caption{HI column density distribution function, $f_{\rm HI}(N_{\rm HI})$, obtained by assigning HI to gas particles employing two different methods: halo based (dotted lines) and particle based (solid lines) described in \ref{sub:bagla} \ref{sub:dave}. Observational data are plotted in black for \cite{Noterdaeme_2012} (crosses) and for \cite{Zafar_2013} (circles). We show results at redshifts $z=3$ (left), $z=4$ (middle) and $z=5$ (right). In each panel we plot the results for the CDM simulation in blue whereas results for the 1 keV WDM cosmology are plotted in red.}
\label{fig:f_HI}
\end{figure}

\begin{figure}
\centering
\includegraphics[width=0.7\textwidth]{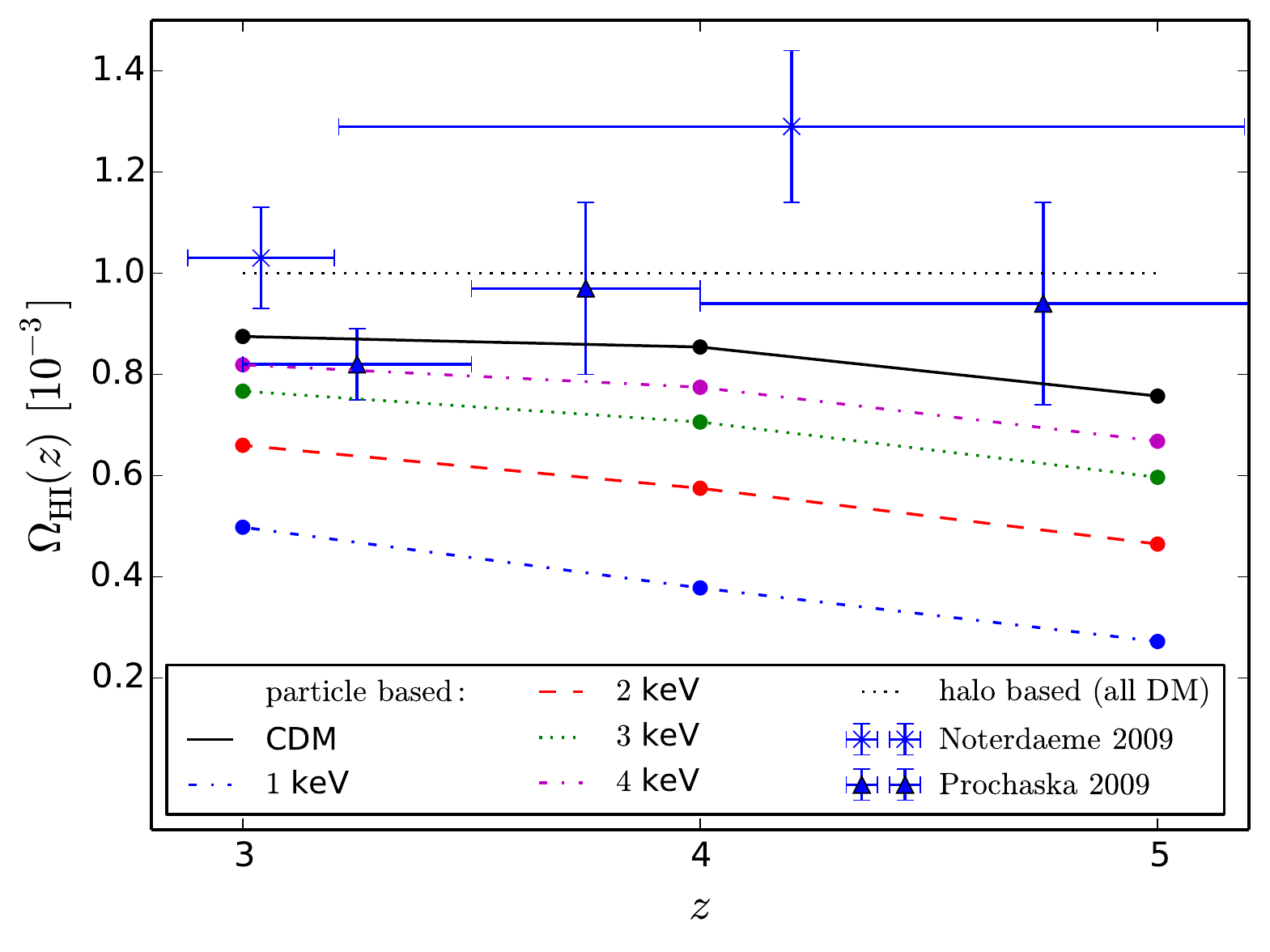}
\caption{Value of the parameter $\Omega_{\rm HI}(z)=\bar{\rho}_{\rm HI}(z)/\rho_c^0$, as a function of redshift, obtained by distributing HI according to the halo based method (dotted line, all cosmologies) and to the particle based method (colored lines). Observational measurements are displayed with error bars in blue, crosses for \cite{Noterdaeme_2009} and triangles for \cite{Prochaska_2009}.}
\label{fig:Omega_HI}
\end{figure}

Up to present days our understanding of the morphology of the hydrogen content in the high redshift universe comes mainly from quasar absorption spectra, where the presence of the Lyman-$\alpha$ transition lines is a conclusive evidence of intervening gas between us and the sources. From these observations we can also infer that most of the hydrogen in the intergalactic medium is ionised, and the neutral hydrogen is mainly contained in the so called damped Lyman-$\alpha$ absorber systems (DLAs) with HI column densities $N_{\rm HI}$ above $2 \times 10^{20}$ atoms cm$^{-2}$ and secondarily in Lyman limit systems (LLS), at lower HI column densities. In order to test the way we distribute HI in the simulations, we compare the abundance of HI absorbers in our simulations with observational data from DLAs \cite{Noterdaeme_2012} and LLS \cite{Zafar_2013}. We have assumed that the HI column density distribution function, from observations, does not vary in the redshift range studied in this paper.

The column density distribution function, $f_{\rm HI}(N_{\rm HI})$, is computed for each simulation, redshift, and for both HI assignment methods by projecting the HI particles position on a plane and picking random lines of sight perpendicular to the plane. For each line the HI column density is computed using the SPH radius and HI mass of all gas particles. Details of the computation can be found in \cite{Villaescusa_2014a}.

Results are shown in figure \ref{fig:f_HI} for both the halo based and particle based method, together with the observational measurements. For clarity, we only show the results for two cosmological models: the models with CDM and 1 keV WDM, since these two are the extreme scenarios and together they span the interesting range of haloes that it is also probed by the other models.
We find that the HI column density distribution obtained by distributing the neutral hydrogen with the two different methods considered here is in overall good agreement with the data.

For the halo based model (dotted lines in figure \ref{fig:f_HI}) the agreement works better for column densities higher than $10^{21}$ cm$^{-2}$ while the abundance of absorbers with lower column densities is overestimated by this method. We find that our results are redshift and cosmology dependent, as the overestimation is weaker at  higher redshifts and for low keV mass WDM, which is in turn due to the lack of low mass halos in WDM scenarios where low column density absorbers reside.

The mismatch in the abundance of LLS between our results and observations does not weaken the results of this work, since their contribution to the total HI distribution is marginal compared to the DLAs, which are instead well reproduced.

By simulating the HI distribution using the particle based method (solid lines in figure \ref{fig:f_HI}), the agreement is better for low HI column density absorbers while the abundance of absorbers with large column densities is underestimated at all redshift and in all cosmologies. 

Generally speaking, the two different HI-assignment methods are somehow complementary: one overestimates what the other underestimates, leaving the data points wrapped by the two approaches. We stress that the halo based and the particle based methods are greatly different both in spirit and in the actual details of the implementation as discussed in Sec. \ref{sec:HI_distribution}, but, looking at the $f_{\rm HI}(N_{\rm HI})$ results, they reasonably well reproduce observations.

\subsection{The HI power spectrum}
\label{sub:Pk_HI}

\begin{figure}
\includegraphics[width=\textwidth]{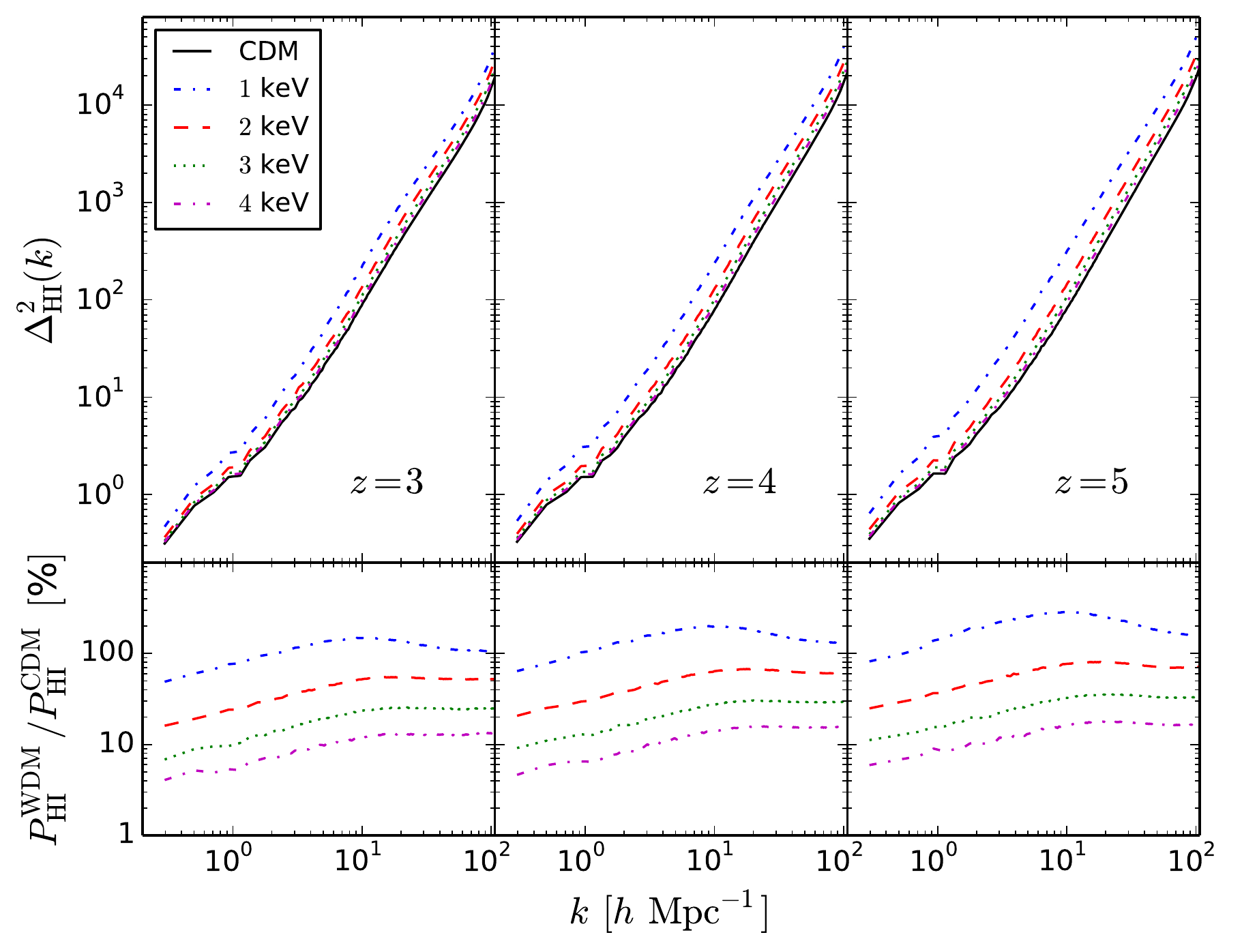}
\caption{Dimensionless HI power spectrum, $\Delta^2_{\rm HI}(k)=k^3 P_{\rm HI}(k) / (2 \pi^2)$,
at redshifts $z=3$ (left), $z=4$ (middle) and $z=5$ (right) obtained by assigning HI to gas particles according to the halo based method. In each panel we plot the results of the CDM simulation with a continuous black line and for the WDM ones: $1$ keV mass (dash dotted blue), $2$ keV (dashed red), $3$ keV (dotted green) and $4$ keV (dash dotted magenta). The relative difference, in percentage, of the WDM models with respect to the CDM scenario is shown in the bottom panels. We set the $k$ range up to  the Nyquist frequency $(k\simeq 107\, h\, {\rm Mpc}^{-1} )$.}
\label{fig:Pk_HI_bagla}
\end{figure}
 
\begin{figure}
\includegraphics[width=\textwidth]{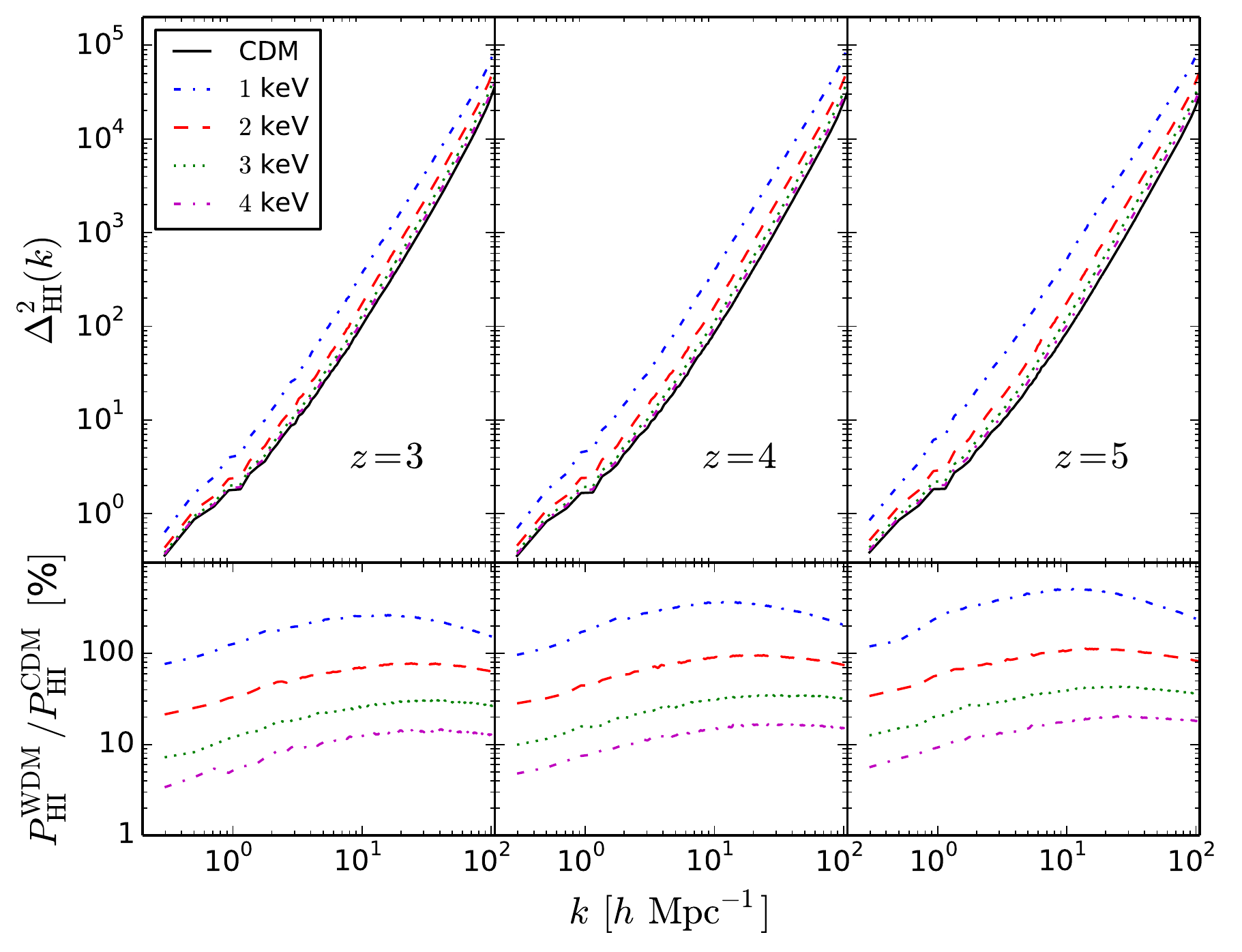}
\caption{Same as figure \ref{fig:Pk_HI_bagla} but modeling the distribution of HI using the particle based method.}

\label{fig:Pk_HI_dave}
\end{figure}

We now investigate the impact of WDM on the spatial distribution of neutral hydrogen. We perform this task by using the simplest statistical estimator that provides information on the spatial distribution of neutral hydrogen: the HI power spectrum.

We display the HI power spectrum, when the HI distribution is modeled using the halo based method, for all the cosmological models studied in this paper in figure \ref{fig:Pk_HI_bagla}, while the results when the particle based method is used are shown in figure \ref{fig:Pk_HI_dave}.
 The five different cosmology power spectra $P_{\rm HI}(k)$ get closer to each other, at all redshifts, on large scales, although they never exactly converge as happen for the total matter power spectra $P_{\rm m}(k)$ showed in figure \ref{fig:matterPk}.
As expected, the lighter the WDM mass, the bigger is the discrepancy with the CDM case, with an increase of power. We find that whereas the HI power spectrum for the 1 keV model has an amplitude roughly  a factor two higher than the one from the CDM model, almost independently on the model used to distribute the HI, differences between the 4 keV and CDM models are smaller than $\sim10\%$. 

In the bottom panels of figures \ref{fig:Pk_HI_bagla} and \ref{fig:Pk_HI_dave}, we plot the ratio between the HI power spectra of the models with WDM to the one of the model with CDM. In the halo based method, the difference in HI power is at least at the $\sim 5\%$ level for the 4 keV WDM cosmology at redshift $z = 3$ and goes up to over the $\sim100\%$ for the 1 keV WDM at $z = 5$. By employing the particle based, the differences are even higher, reaching more than $300\%$ in the most extreme case of 1 keV at $z=5$.

\begin{figure}
\includegraphics[width=\textwidth]{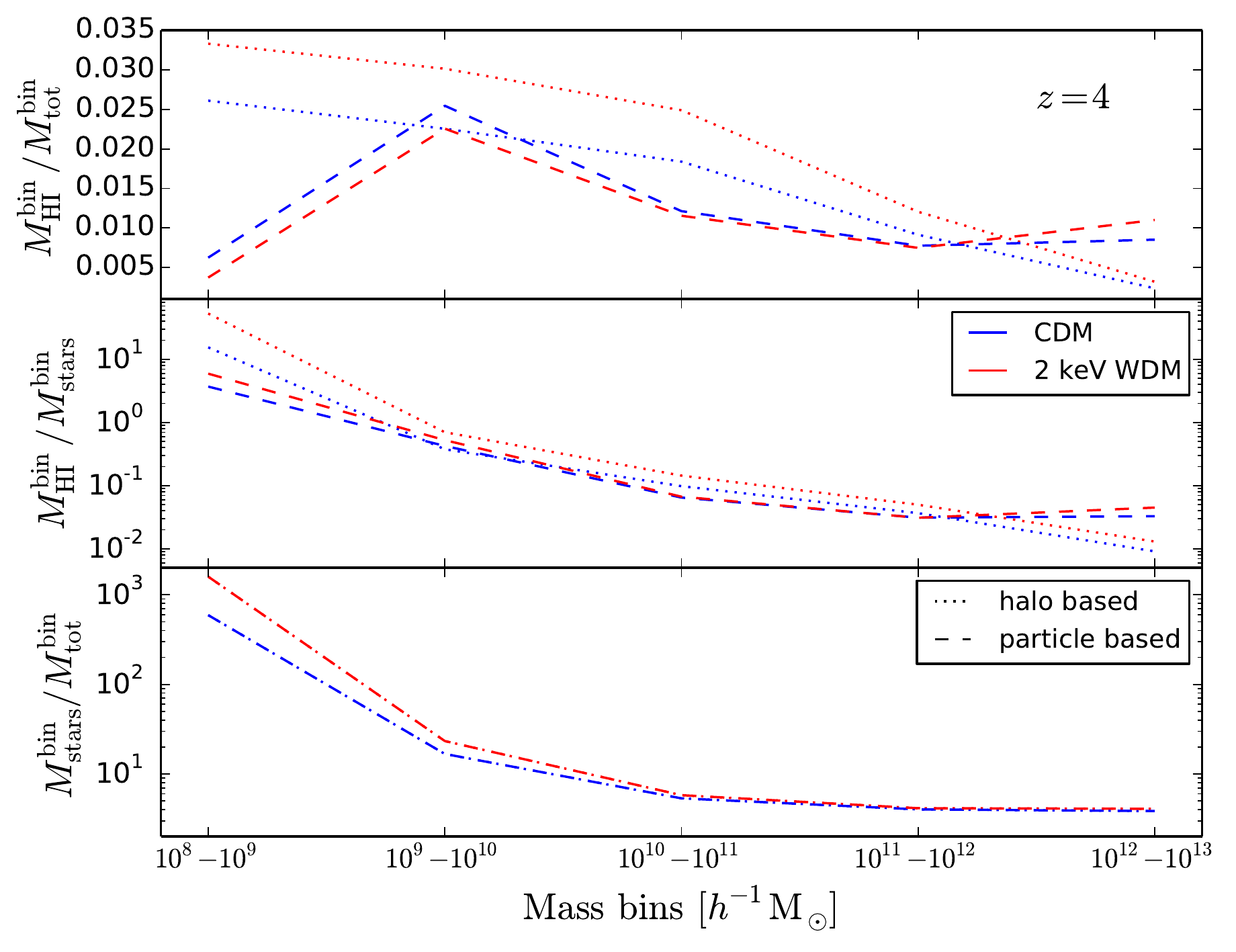}
\caption{Ratios of the total HI mass over the matter $M_{\rm HI}/M_{\rm tot}$ (upper panel), over the stellar mass $M_{\rm HI}/M_{\rm stars}$ (middle panel) and the total stellar mass over the total mattel $M_{\rm stars}/M_{\rm tot}$ (lower panel) calculated as function of the halo mass bins within the ratio is performed, for the CDM (blue lines) and 2 keV WDM (red lines) cosmologies with the halo based (dotted lines) and particle based (dashed lines) HI assignment methods at redshift $z=4$.}
\label{fig:mass_bins}
\end{figure}

The reason why the amplitude of the HI power spectrum is higher in the models with WDM with respect to the CDM model is straightforward when the neutral hydrogen distribution is modeled using the halo based method. By using this method we are forced to distribute the same amount of HI in the cosmologies with CDM and WDM. Since we assume that halos with masses smaller than $M_{\rm min}$ do not host HI, and since there is a deficit of low-mass halos in the cosmologies with WDM (in comparison to CDM), we need to put more HI into the remaining halos when we assign HI to the halos of the WDM cosmologies. Thus, the HI will be more clustered in the models with WDM because we have less halos at disposal to be filled in and those available have higher masses and therefore larger bias.

When the HI is modeled using the particle based method it is not that obvious the reason of the larger clustering of the neutral hydrogen in the WDM cosmologies.
To further investigate this issue we have performed the following test. We select all dark matter halos in a given mass interval and compute the sum of the HI mass, stellar mass and total mass within those halos for the models with CDM and 2 keV WDM. In figure \ref{fig:mass_bins} we plot the ratio between the total HI mass and the total mass contained in the halos in the selected mass range (top panel), the ratio between the total HI and stellar mass (middle panel) and the ratio between the total stellar and total mass (bottom panel). 

Looking at the ratio $M_{\rm HI}^{\rm bin}/M_{\rm tot}^{\rm bin}$ in the top panel, we see the cutoff of the HI in the high mass halos predicted by the halo based model, whereas the particle based results do not show such decline. From that panel it is also clear that halos of the same mass have more HI in the models with WDM with respect to the CDM model, when the HI is modeled using the halo based method. $M_{\rm HI}^{\rm bin}/M_{\rm tot}^{\rm bin}$ is very similar in the two different models when the HI is modeled using the particle based method. These results point out that the amount of HI per total mass in a given dark matter halo is almost the same between halos in CDM and WDM cosmologies. Thus, the larger amplitude of the HI power spectrum in the models with WDM arises because the bias of the dark matter halos is higher in the WDM models.

The ratio $M_{\rm HI}^{\rm bin}/M_{\rm stars}^{\rm bin}$ in the middle panel of figure \ref{fig:mass_bins} is higher for WDM and comparable among the halo and particle based methods for the central mass range bins. The ratio of the stellar mass over total mass $M_{\rm stars}^{\rm bin}/M_{\rm tot}^{\rm bin}$ in the lower panel of course does not change with the different HI assignment methods, and it is higher for WDM: it could be sign of more active star formation in the WDM halos (but see for example \cite{maio15}).

In the next section we analyse in more detail the different bias that HI exhibit in CDM and WDM cosmologies, to check whether the total matter clustering properties are indeed reflected by the HI clustering properties.

\subsection{The HI bias}
\label{sub:bias_HI}

\begin{figure}
\includegraphics[width=\textwidth]{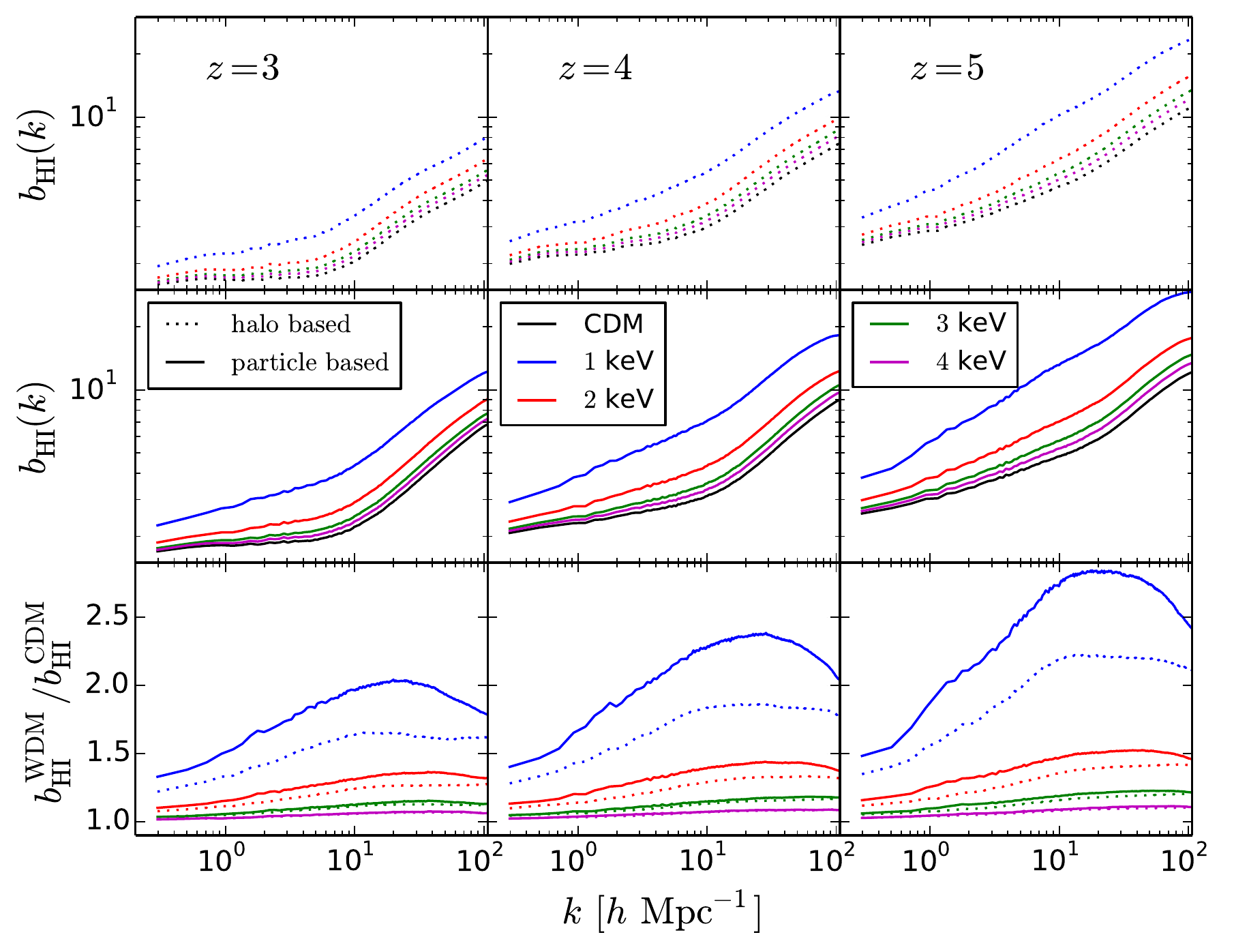}
\caption{HI bias, $b_{\rm HI}(k)=\sqrt{P_{\rm HI}(k)/P_{\rm m}(k)}$, at $z=3$ (left column), $z=4$ (middle column) and $z=5$ (right column) when the HI distribution is modeled using the halo based (top row) and the particle based (middle row) methods. In each panel we plot the results for the CDM model with black lines whereas we show with colored lines the results for the models with WDM: $1$ keV (blue), $2$ keV (red), $3$ keV (green) and $4$ keV (magenta). The ratio between the bias of the models with WDM to the
model with CDM is displayed in the bottom row. We set the $k$ range up to  the Nyquist frequency $(k\simeq 107\, h\, {\rm Mpc}^{-1} )$.}
\label{fig:bias_HI}
\end{figure}

After having computed the total matter and the neutral hydrogen power spectra $P_{\rm m}(k)$ and ${P_{\rm HI}(k)}$, respectively, we can now estimate the HI bias, $b_{\rm HI}(k)$, defined as:
\be
b_{\rm HI}^2(k) = \f{P_{\rm HI}(k)}{P_{\rm m}(k)}\,.
\ee

In figure \ref{fig:bias_HI} we plot the HI bias at redshifts $z=3$ (left), $z=4$ (middle) and $z=5$ (right) when the HI distribution is modeled using the halo based (top row) and the particle based (middle row) methods. We find that the spatial distribution of neutral hydrogen is more biased in the WDM models, with the bias increasing with decreasing WDM mass, at all redshift, for both methods. Among the two different models, the particle based produces a slightly higher HI bias in comparison to the results obtained by employing the halo based model. This can be more easily seen in the bottom panels of figure \ref{fig:bias_HI}, where we plot the relative difference in the bias between the models with WDM and CDM, for both methods. As discussed in the previous section, the reason why the HI bias is higher in the models with WDM is because the neutral hydrogen is more strongly clustered in these models in comparison to model with CDM.

\subsection{The 21cm power spectrum and SKA1-LOW  forecasts}
\label{sub:21Pk}

Radio telescopes can detect the redshifted 21cm radiation from neutral hydrogen. Therefore, the quantity directly measured from observations is not the HI power spectrum, but the 21cm power spectrum, which is nothing but the power spectrum of the spatial distribution of neutral hydrogen in redshift-space. For concreteness, the 21cm power spectrum is defined as
\begin{equation}
P_{\rm 21cm}(k)=\langle \delta T_b(\vec{k}) (\delta T_b)^*(\vec{k})  \rangle,
\label{Pk_21cm_s}
\end{equation}
where $\delta T_b$ is given by
\begin{equation}
\delta T_b(\nu)=\overline{\delta T_b}(z)\left[\frac{\rho_{\rm HI}(\vec{s})}{\bar{\rho}_{\rm HI}}\right],
\label{delta_Tb}
\end{equation}
with $\rho_{\rm HI}(\vec{s})$ being the neutral hydrogen density in the redshift-space position $\vec{s}$ and 
\begin{equation}
\overline{\delta T_b}(z)=23.88~\bar{x}_{\rm HI}\left( \frac{\Omega_{\rm b}h^2}{0.02}\right)\sqrt{\frac{0.15}{\Omega_{\rm m}h^2}\frac{(1+z)}{10}}~{\rm mK},
\end{equation}
where $\bar{x}_{\rm HI}=\bar{\rho}_{\rm HI}/\bar{\rho}_{\rm H}$ is the average neutral hydrogen fraction. Here we compute the 21cm power spectrum for the different cosmological models investigated in this paper and study the sensitivity with which the future SKA1-LOW array will be able to discriminate among different models.

Starting from the spatial distribution of neutral hydrogen in real-space, that we obtain using the two methods described in Sec. \ref{sec:HI_distribution}, we then displace the particle positions to obtain the coordinates of the particles in redshift-space by doing
\be
\vec{s}=\vec{x}+\frac{1+z}{H(z)} \vec{v}_{\rm los}(\vec{x})\,,
\ee
with $z$ being the redshift of observation, $\vec{v}_{\rm los}$ the line of sight component of the peculiar velocity and $H(z)$ the Hubble parameter. 
For convenience, we prefer to work with the dimensionless 21cm power spectrum $\Delta_{21\rm cm}^2(k)$, namely:
\be
\Delta_{21\rm cm}^2(k)=\frac{k^3 P_{21\rm cm}(k)}{2 \pi^2} \,.
\ee

We have computed the 21cm power spectrum for each cosmological model and for each redshift using each of the two methods employed to model the distribution of HI. For the halo based method  $\Omega_{\rm HI}^{\rm ref} = 10^{-3}$ is fixed in all scenarios by construction, whereas for the particle based, we have different values of $\Omega_{\rm HI}^{\rm sim}$ for each simulation.

Since we want to investigate differences between models, and we assume that the value of $\Omega_{\rm HI}$ is fixed by independent observations (such as the abundance of DLAs and LLS), a further normalization is needed when computing the 21cm power spectrum, to force all models to have the same value of $\Omega_{\rm HI}$\footnote{Notice that by doing this we are assuming that the clustering properties of the neutral hydrogen do not change.}:
\be
(\Delta_{21\rm cm}^2(k) )^{\rm norm} = \Delta_{21\rm cm}^2(k) \left(\frac{\Omega_{\rm HI}^{\rm ref}}{\Omega_{\rm HI}^{\rm sim}}\right)^2 \,, 
\label{eq:Pk_norm}
\ee
where $\Omega_{\rm HI}^{\rm sim}$ is the value of $\Omega_{\rm HI}$ directly obtained by employing the particle based method.

\begin{figure}
\includegraphics[width=\textwidth]{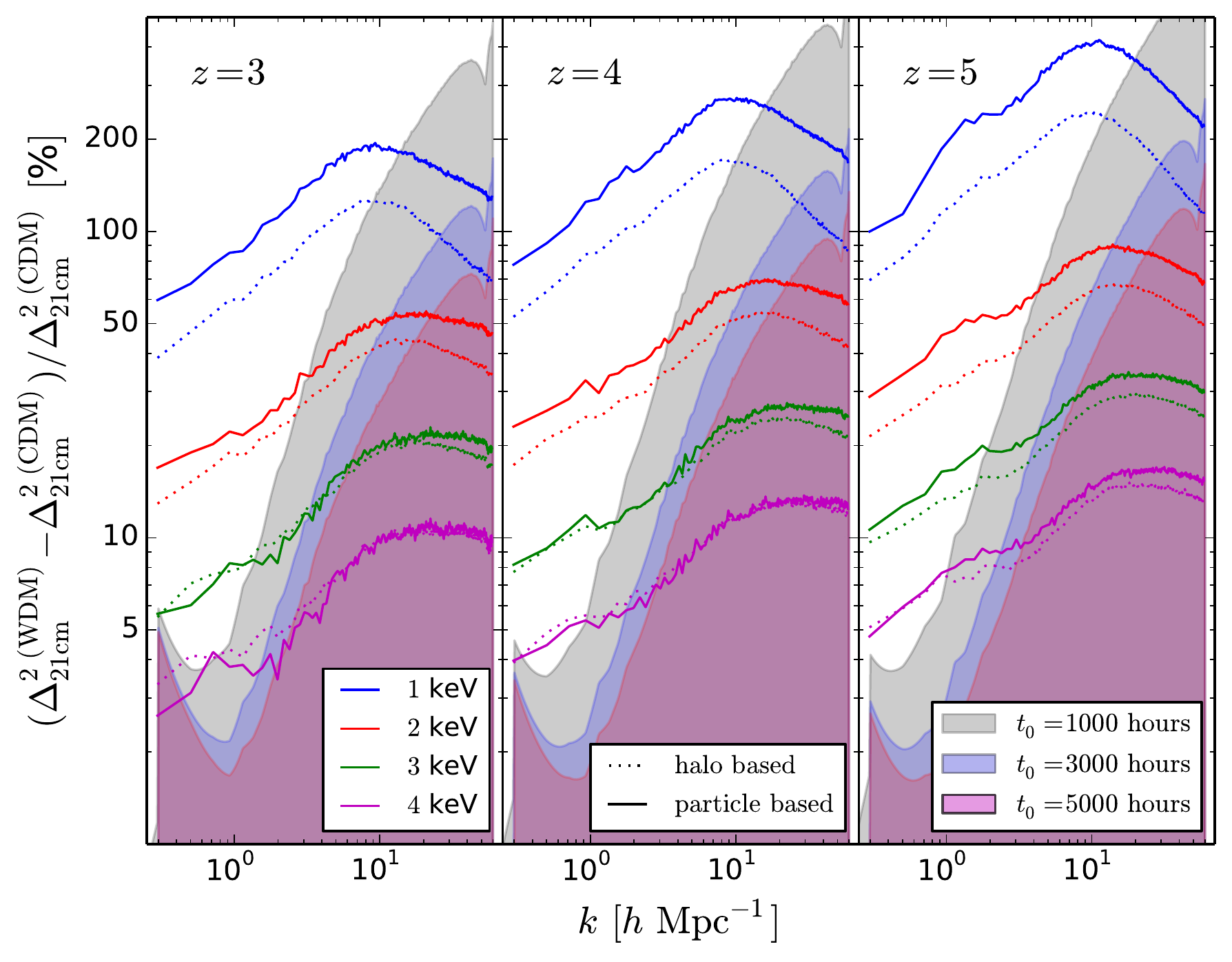}
\caption{Relative difference between the 21cm power spectrum of the models with WDM and CDM when the HI distribution is modeled using the halo based method (dotted lines) and the particle based method (solid lines). Results are shown at $z=3$ (left), $z=4$ (middle) and $z=5$ (right). The error on the 21cm power spectrum of the model with CDM, normalized to the amplitude of the 21cm power spectrum, $\sigma \Delta_{21\rm cm}^{2\,{\rm (CDM)}}/\Delta_{21\rm cm}^{2\,{\rm (CDM)}}$, is shown in a shaded region for three different observation times: $t_0=1000$ hours (grey), $t_0=3000$ hours (blue) and $t_0=5000$ hours (fuchsia). For clarity, we show the error on $\Delta_{21\rm cm}^{2\,{\rm (CDM)}}$ from one HI-assignment method only because both are vey similar and overlap at the scale of the plot.}
\label{fig:21cmPk}
\end{figure}

In figure \ref{fig:21cmPk} we show the relative difference in the 21cm power spectrum between the models with WDM and the model with CDM. Dotted lines represents the results when the halo based model is used whereas solid lines display the results when the particle based model is employed. 

As with the HI power spectrum, we find that the amplitude of the 21cm power spectrum is higher in the models with WDM than in the model with CDM. Again, this is because the spatial distribution of neutral hydrogen is more strongly clustered in the models with WDM. We have also computed the errors with which the SKA1-LOW radio-telescope will measure the 21cm power spectrum. Details of the error computation can be found in \cite{Villaescusa_2014b}. The shaded areas in figure \ref{fig:21cmPk} show the quantity $\sigma \Delta_{21\rm cm}^{2\,{\rm (CDM)}}/\Delta_{21\rm cm}^{2\,{\rm (CDM)}}$, where $\sigma \Delta_{21\rm cm}^{2\,{\rm (CDM)}}$ represents the $1\sigma$ error on the 21cm dimensionless power spectrum of the model with CDM. Results are shown for observations times equal to $t_0=1000$ hours (grey), $t_0=3000$ hours (blue) and $t_0=5000$ hours (fuchsia). 

We find that with a reasonable observational time of $t_0=1000$ hours \cite{Pritchard_2015} the WDM models with 1,2, and 3 keV can be distinguished from  CDM. On the other hand, the model with 4 keV WDM is consistent with CDM at the $\sim 1\sigma$ confidence level at $z=5$. With longer observational times, e.g. 3000 hours, even the model with 4 keV can be distinguished from the model with CDM at more than $2\sigma$ at $z=5$. We note that on large scales, the error budget is dominated by cosmic variance, thus, the error magnitude barely changes by increasing the observational time. Extending the observation time to $t_0=5000$ hours, also the 4 keV case is above the error at  $\sim 2.5 \sigma$ confidence level for 21cm emission coming from redshift $z=3$. These conclusions are valid for both the methods considered in this paper.

\section{Summary and conclusions}
\label{sec:conclusions}

The standard $\Lambda$CDM cosmological model, which has been recently tested with unprecedented accuracy by Planck \cite{Planck_2015}, has been proved to be extremely successful in reproducing large scale structure observables from the cosmic microwave background to the Lyman-$\alpha$ forest. However, some properties of the matter distribution at very small scales ($< 1~h^{-1}$ Mpc) are difficult to reconcile within this model. While this putative crisis could be solved by baryonic physics (i.e. feedback), it is intriguing
to explore other possibilities, like the fact that thermal velocities of the dark matter component could affect  the properties of matter at small scales (see e.g. the discussion in \cite{weinberg} where other modifications of the CDM paradigm are discussed for lifting/solving these problems).
The matter power spectrum in a cosmology with WDM presents a cut-off on small scales, that arises because WDM can not significantly cluster on scales smaller than their mean free path. The tightest constrains on the mass of the WDM comes from observations of the Lyman-$\alpha$ forest, which are used to constrain the shape of the matter power spectrum on small scales: $m_{\rm WDM}\geqslant3.3~{\rm keV}(2\sigma)$ \cite{Viel_2013}.

In this work we focus on a relatively new observable: intensity mapping. The idea is to perform a low angular resolution survey to measure the integrated redshifted 21cm emission from unresolved galaxies (see for instance \cite{Bull_2014}) which will allow to measure the spatial distribution of neutral hydrogen, in terms of the 21cm power spectrum. 

In order to do that we investigate the spatial distribution of neutral hydrogen in cosmologies with WDM and study the signatures left by those on the shape and amplitude of the 21cm power spectrum. We also forecast which with sensitivity will the SKA1-LOW array be able to distinguish between different cosmological models.

Our assumptions rely on the fact that the cosmological parameters are fixed to exquisite precision by cosmic microwave background experiments,  and a further parameter is added that quantifies the suppression of power: we do not explore scenarios in which feedback is important and we assume a reference thermal history that sets the physical properties of the gas in the post-reionization era. On one side,  our analysis is aggressive in the sense that a more comprehensive analysis of the expected signal should probably account for different astrophysical processes (e.g. radiative transfer effects) that could be modeled and marginalized over to quote the final level of significance of a WDM model. On the other side, we want to be conservative in two ways: i) we decide to focus on a regime which is far from the complex astrophysics of reionization and possibly also not affected at an appreciable level by galactic feedback; ii) we use two different models to distribute HI in the Universe. Furthermore, this regime has the advantage of being constrained by data (the total amount of HI from absorption lines).

Our simulation suite comprises a set of 5 high-resolution hydrodynamic  simulations. We simulate a cosmological model with CDM and 4 models with WDM with different particle masses: 1, 2, 3 and 4 keV. We model the spatial distribution of neutral hydrogen by assigning, a-posteriori, HI to the gas particles in the simulations, by using: i) a halo-based model, in which HI is assigned only to gas particles belonging to dark matter halos and ii) a particle-based method in which HI is assigned to every single gas particle in the simulation according to its physical properties.
We  compute the HI column density distribution function (CDDF) at redshifts $z=3,4,5$ and find good agreement with observational data in all cases, although models with low WDM masses fail at reproducing the HI CDDF for absorbers with low column density when the particle-based model is employed. All models predict a lower abundance of absorbers with large column densities by using the particle-based method. By distributing the HI according to the halo-based model we find an excellent agreement with observations for all models, except for the LLS, whose abundance is over predicted by using this method.

We then extract the HI power spectrum for the cosmologies with CDM and WDM. 
Contrary to a naive expectation, for which a similar suppression of power should be found in the HI and 21cm power spectra  we find instead that the amplitude of the HI power spectrum is higher in the models with WDM (and increases with decreasing WDM mass) than in the model with CDM, at all redshifts studied in this paper and independently of the method used to model the distribution of HI. The reason is that in the WDM cosmologies we have a deficit of low-mass halos, in comparison with the CDM model (see figure \ref{fig:MF}), which means that in the WDM cosmologies the fraction HI residing in the most massive halos is higher in the WDM cosmologies. Therefore, the bias of the HI, $b^2_{\rm HI}(k)=P_{\rm HI}(k)/P_{\rm m}(k)$, will increase with decreasing WDM mass.
This is also in agreement with what shown in \cite{Kang_2013}, where they infer that WDM low mass halos should exhibit higher baryonic mass compared to the CDM case in order to reproduce the local stellar mass function.

The most interesting conclusion is that the HI power spectrum of the WDM does not have an exponential cut-off, at least on the scales probes in this paper ($\sim 100~h{\rm Mpc}^{-1}$), that it is however present in the matter power spectrum. The reason is that neutral hydrogen found resides in halos of masses larger than $\sim10^{9}~h^{-1}M_\odot$ and the effect of WDM on those halos is to suppress their abundance and thus results in an increased bias. The effect of WDM on the matter power spectrum on small scales is more pronounced since the power on these scales arises mainly from  low mass halos.

Finally, we have computed the 21cm power spectrum for each of our 5 different cosmological models. As in the case of the HI power spectrum, we find that the amplitude of the 21cm power spectrum is higher in the models with WDM than in the models with CDM. We find that the amplitude of the 21cm power spectrum is up to $400 \%$ higher for the 1 keV case at redshift $z=5$ and down to $10 \%$ higher for the 4 keV at redshift $z=3$, when comparing it with the 21cm power spectrum of the CDM model.

We have also forecasted the observing time required for the SKA1-LOW array to be able to distinguish between models with WDM and CDM. Since differences between those models increase with redshift, we find that models are more easily distinguishable at $z=5$ rather than at $z=3$ independently of the method used to model the distribution of HI. Our results point out that with 1000 hours of observations a WDM model with 3 keV can be ruled out at more than $2\sigma$ at $z=5$, while SKA1-LOW will not be able to discriminate among these two models at redshift $z=3$. We find that with 5000 hours of observations the 4 keV WDM model can be ruled out at more than $1\sigma$ at $z=3$ and at more than $2\sigma$ at $z=5$.
In deriving these numbers above, we made use only of the largest scales $k<1-3~h~{\rm Mpc}^{-1}$ available, since the small scale signal is hindered by noise (see figure \ref{fig:21cmPk}) and thereby feedback effects, if present, should affect the signal to a very small extent in these range of scales and redshifts.

We stress that in \cite{sitwell} authors investigated the signatures of WDM in the 21cm power spectrum during the epoch of reionization. For WDM models with masses of 2 and 4 keV they found differences in the 21cm power spectrum as large as 30\% (for the 4 keV model) and 240\% (for the 2 keV model), depending on redshift and scale. These differences are much higher that the ones we find in this paper since differences between CDM and WDM models increase with redshift. On the other hand, in the epoch of reionization the 21cm power spectrum is much severely affected by astrophysical processes, which should be less important once reionization is over. Thus, observations during and after reionization are complementary and will increase the robustness of the constrains on the WDM models.

Regardless of the results obtained for the SKA1-LOW telescope, our framework explores a possible new way to interpret the clustering of matter at small scales and to constrain the fundamental nature of dark matter and thereby strengthen the scientific case for conducting HI/21cm surveys in the high redshift universe.

\section*{Acknowledgements}
N-body simulations were run on the COSMOS Consortium supercomputer 
within the DiRAC  Facility jointly funded by STFC, the Large Facilities Capital 
Fund of BIS and the University of Cambridge, as well as the Darwin 
Supercomputer of the University of Cambridge High Performance Computing 
Service (http://www.hpc.cam.ac.uk/), provided by Dell Inc. using Strategic Research
Infrastructure Funding from the Higher Education Funding Council for
England.  FVN and MV are supported by the ERC Starting Grant
``cosmoIGM'' and partially supported by INFN IS PD51 "INDARK". MV is also supported
by PRIN MIUR 2012 and PRIN INAF  2011 "A complete view of the first 2 billion years of galaxy formation".
We acknowledge partial support from ``Consorzio per la Fisica -- Trieste''.


\bibliographystyle{JHEP}
\bibliography{Bibliography} 

\providecommand{\href}[2]{#2}\begingroup\raggedright\begin{thebibliography}{10}

\bibitem{Planck_2015}
{Planck Collaboration}, P.~A.~R. {Ade}, N.~{Aghanim}, M.~{Arnaud},
  M.~{Ashdown}, J.~{Aumont}, C.~{Baccigalupi}, A.~J. {Banday}, R.~B.
  {Barreiro}, J.~G. {Bartlett}, and et~al., {\it {Planck 2015 results. XIII.
  Cosmological parameters}},  {\em ArXiv e-prints} (Feb., 2015)
  [\href{http://xxx.lanl.gov/abs/1502.0158}{{\tt arXiv:1502.0158}}].

\bibitem{NFW}
J.~F. {Navarro}, C.~S. {Frenk}, and S.~D.~M. {White}, {\it {A Universal Density
  Profile from Hierarchical Clustering}},  {\em \apj} {\bf 490} (Dec., 1997)
  493--508, [\href{http://xxx.lanl.gov/abs/astro-ph/9611107}{{\tt
  astro-ph/9611107}}].

\bibitem{Aquarius}
V.~{Springel}, J.~{Wang}, M.~{Vogelsberger}, A.~{Ludlow}, A.~{Jenkins},
  A.~{Helmi}, J.~F. {Navarro}, C.~S. {Frenk}, and S.~D.~M. {White}, {\it {The
  Aquarius Project: the subhaloes of galactic haloes}},  {\em \mnras} {\bf 391}
  (Dec., 2008) 1685--1711, [\href{http://xxx.lanl.gov/abs/0809.0898}{{\tt
  arXiv:0809.0898}}].

\bibitem{Salucci_2007}
P.~{Salucci}, A.~{Lapi}, C.~{Tonini}, G.~{Gentile}, I.~{Yegorova}, and
  U.~{Klein}, {\it {The universal rotation curve of spiral galaxies - II. The
  dark matter distribution out to the virial radius}},  {\em \mnras} {\bf 378}
  (June, 2007) 41--47, [\href{http://xxx.lanl.gov/abs/astro-ph/0703115}{{\tt
  astro-ph/0703115}}].

\bibitem{Gilmore_2007}
G.~{Gilmore}, M.~I. {Wilkinson}, R.~F.~G. {Wyse}, J.~T. {Kleyna}, A.~{Koch},
  N.~W. {Evans}, and E.~K. {Grebel}, {\it {The Observed Properties of Dark
  Matter on Small Spatial Scales}},  {\em \apj} {\bf 663} (July, 2007)
  948--959, [\href{http://xxx.lanl.gov/abs/astro-ph/0703308}{{\tt
  astro-ph/0703308}}].

\bibitem{Eymeren_2009}
J.~{van Eymeren}, C.~{Trachternach}, B.~S. {Koribalski}, and R.-J. {Dettmar},
  {\it {Non-circular motions and the cusp-core discrepancy in dwarf galaxies}},
   {\em \aap} {\bf 505} (Oct., 2009) 1--20,
  [\href{http://xxx.lanl.gov/abs/0906.4654}{{\tt arXiv:0906.4654}}].

\bibitem{Naray_2010}
R.~{Kuzio de Naray}, G.~D. {Martinez}, J.~S. {Bullock}, and M.~{Kaplinghat},
  {\it {The Case Against Warm or Self-Interacting Dark Matter as Explanations
  for Cores in Low Surface Brightness Galaxies}},  {\em \apjl} {\bf 710} (Feb.,
  2010) L161--L166, [\href{http://xxx.lanl.gov/abs/0912.3518}{{\tt
  arXiv:0912.3518}}].

\bibitem{Walker_2011}
M.~G. {Walker} and J.~{Pe{\~n}arrubia}, {\it {A Method for Measuring (Slopes
  of) the Mass Profiles of Dwarf Spheroidal Galaxies}},  {\em \apj} {\bf 742}
  (Nov., 2011) 20, [\href{http://xxx.lanl.gov/abs/1108.2404}{{\tt
  arXiv:1108.2404}}].

\bibitem{weinberg}
D.~H. {Weinberg}, J.~S. {Bullock}, F.~{Governato}, R.~{Kuzio de Naray}, and
  A.~H.~G. {Peter}, {\it {Cold dark matter: controversies on small scales}},
  {\em ArXiv e-prints} (June, 2013)
  [\href{http://xxx.lanl.gov/abs/1306.0913}{{\tt arXiv:1306.0913}}].

\bibitem{ElZant_2001}
A.~{El-Zant}, I.~{Shlosman}, and Y.~{Hoffman}, {\it {Dark Halos: The Flattening
  of the Density Cusp by Dynamical Friction}},  {\em \apj} {\bf 560} (Oct.,
  2001) 636--643, [\href{http://xxx.lanl.gov/abs/astro-ph/0103386}{{\tt
  astro-ph/0103386}}].

\bibitem{Tonini_2006}
C.~{Tonini}, A.~{Lapi}, and P.~{Salucci}, {\it {Angular Momentum Transfer in
  Dark Matter Halos: Erasing the Cusp}},  {\em \apj} {\bf 649} (Oct., 2006)
  591--598, [\href{http://xxx.lanl.gov/abs/astro-ph/0603051}{{\tt
  astro-ph/0603051}}].

\bibitem{Maccio_2012b}
A.~V. {Macci{\`o}}, G.~{Stinson}, C.~B. {Brook}, J.~{Wadsley}, H.~M.~P.
  {Couchman}, S.~{Shen}, B.~K. {Gibson}, and T.~{Quinn}, {\it {Halo Expansion
  in Cosmological Hydro Simulations: Toward a Baryonic Solution of the
  Cusp/Core Problem in Massive Spirals}},  {\em \apjl} {\bf 744} (Jan., 2012)
  L9, [\href{http://xxx.lanl.gov/abs/1111.5620}{{\tt arXiv:1111.5620}}].

\bibitem{Schneider_2014}
A.~{Schneider}, D.~{Anderhalden}, A.~V. {Macci{\`o}}, and J.~{Diemand}, {\it
  {Warm dark matter does not do better than cold dark matter in solving
  small-scale inconsistencies}},  {\em \mnras} {\bf 441} (June, 2014) L6--L10,
  [\href{http://xxx.lanl.gov/abs/1309.5960}{{\tt arXiv:1309.5960}}].

\bibitem{Maccio_2013}
A.~V. {Macci{\`o}}, O.~{Ruchayskiy}, A.~{Boyarsky}, and J.~C.
  {Mu{\~n}oz-Cuartas}, {\it {The inner structure of haloes in cold+warm dark
  matter models}},  {\em \mnras} {\bf 428} (Jan., 2013) 882--890,
  [\href{http://xxx.lanl.gov/abs/1202.2858}{{\tt arXiv:1202.2858}}].

\bibitem{Villaescusa-Navarro_Dalal}
F.~{Villaescusa-Navarro} and N.~{Dalal}, {\it {Cores and cusps in warm dark
  matter halos}},  {\em \jcap} {\bf 3} (Mar., 2011) 24,
  [\href{http://xxx.lanl.gov/abs/1010.3008}{{\tt arXiv:1010.3008}}].

\bibitem{Maccio_2012}
A.~V. {Macci{\`o}}, S.~{Paduroiu}, D.~{Anderhalden}, A.~{Schneider}, and
  B.~{Moore}, {\it {Cores in warm dark matter haloes: a Catch 22 problem}},
  {\em \mnras} {\bf 424} (Aug., 2012) 1105--1112,
  [\href{http://xxx.lanl.gov/abs/1202.1282}{{\tt arXiv:1202.1282}}].

\bibitem{Shao_2013}
S.~{Shao}, L.~{Gao}, T.~{Theuns}, and C.~S. {Frenk}, {\it {The phase-space
  density of fermionic dark matter haloes}},  {\em \mnras} {\bf 430} (Apr.,
  2013) 2346--2357, [\href{http://xxx.lanl.gov/abs/1209.5563}{{\tt
  arXiv:1209.5563}}].

\bibitem{Lovell_2014}
M.~R. {Lovell}, C.~S. {Frenk}, V.~R. {Eke}, A.~{Jenkins}, L.~{Gao}, and
  T.~{Theuns}, {\it {The properties of warm dark matter haloes}},  {\em \mnras}
  {\bf 439} (Mar., 2014) 300--317,
  [\href{http://xxx.lanl.gov/abs/1308.1399}{{\tt arXiv:1308.1399}}].

\bibitem{Destri_2014}
C.~{Destri}, {\it {Hollow cores in Warm Dark Matter halos from the
  Vlasov-Poisson equation}},  {\em ArXiv e-prints} (Sept., 2014)
  [\href{http://xxx.lanl.gov/abs/1409.6244}{{\tt arXiv:1409.6244}}].

\bibitem{Bode_2001}
P.~{Bode}, J.~P. {Ostriker}, and N.~{Turok}, {\it {Halo Formation in Warm Dark
  Matter Models}},  {\em \apj} {\bf 556} (July, 2001) 93--107,
  [\href{http://xxx.lanl.gov/abs/astro-ph/0010389}{{\tt astro-ph/0010389}}].

\bibitem{Avila-Reese_2001}
V.~{Avila-Reese}, P.~{Col{\'{\i}}n}, O.~{Valenzuela}, E.~{D'Onghia}, and
  C.~{Firmani}, {\it {Formation and Structure of Halos in a Warm Dark Matter
  Cosmology}},  {\em \apj} {\bf 559} (Oct., 2001) 516--530,
  [\href{http://xxx.lanl.gov/abs/astro-ph/0010525}{{\tt astro-ph/0010525}}].

\bibitem{Schneider_2012}
A.~{Schneider}, R.~E. {Smith}, A.~V. {Macci{\`o}}, and B.~{Moore}, {\it
  {Non-linear evolution of cosmological structures in warm dark matter
  models}},  {\em \mnras} {\bf 424} (July, 2012) 684--698,
  [\href{http://xxx.lanl.gov/abs/1112.0330}{{\tt arXiv:1112.0330}}].

\bibitem{Viel_2013}
M.~{Viel}, G.~D. {Becker}, J.~S. {Bolton}, and M.~G. {Haehnelt}, {\it {Warm
  dark matter as a solution to the small scale crisis: New constraints from
  high redshift Lyman-{$\alpha$} forest data}},  {\em \prd} {\bf 88} (Aug.,
  2013) 043502, [\href{http://xxx.lanl.gov/abs/1306.2314}{{\tt
  arXiv:1306.2314}}].

\bibitem{Bull_2014}
P.~{Bull}, P.~G. {Ferreira}, P.~{Patel}, and M.~G. {Santos}, {\it {Late-time
  cosmology with 21cm intensity mapping experiments}},  {\em ArXiv e-prints}
  (May, 2014) [\href{http://xxx.lanl.gov/abs/1405.1452}{{\tt
  arXiv:1405.1452}}].

\bibitem{Camera_2013}
S.~{Camera}, M.~G. {Santos}, P.~G. {Ferreira}, and L.~{Ferramacho}, {\it
  {Cosmology on Ultralarge Scales with Intensity Mapping of the Neutral
  Hydrogen 21 cm Emission: Limits on Primordial Non-Gaussianity}},  {\em
  Physical Review Letters} {\bf 111} (Oct., 2013) 171302,
  [\href{http://xxx.lanl.gov/abs/1305.6928}{{\tt arXiv:1305.6928}}].

\bibitem{sitwell}
M.~{Sitwell}, A.~{Mesinger}, Y.-Z. {Ma}, and K.~{Sigurdson}, {\it {The imprint
  of warm dark matter on the cosmological 21-cm signal}},  {\em \mnras} {\bf
  438} (Mar., 2014) 2664--2671, [\href{http://xxx.lanl.gov/abs/1310.0029}{{\tt
  arXiv:1310.0029}}].

\bibitem{dayal}
P.~{Dayal}, A.~{Mesinger}, and F.~{Pacucci}, {\it {Early galaxy formation in
  warm dark matter cosmologies}},  {\em ArXiv e-prints} (Aug., 2014)
  [\href{http://xxx.lanl.gov/abs/1408.1102}{{\tt arXiv:1408.1102}}].

\bibitem{maio15}
U.~{Maio} and M.~{Viel}, {\it {The first billion years of a warm dark matter
  universe}},  {\em \mnras} {\bf 446} (Jan., 2015) 2760--2775,
  [\href{http://xxx.lanl.gov/abs/1409.6718}{{\tt arXiv:1409.6718}}].

\bibitem{Evoli_2014}
C.~{Evoli}, A.~{Mesinger}, and A.~{Ferrara}, {\it {Unveiling the nature of dark
  matter with high redshift 21 cm line experiments}},  {\em \jcap} {\bf 11}
  (Nov., 2014) 24, [\href{http://xxx.lanl.gov/abs/1408.1109}{{\tt
  arXiv:1408.1109}}].

\bibitem{Springel_2005}
V.~{Springel}, {\it {The cosmological simulation code GADGET-2}},  {\em \mnras}
  {\bf 364} (Dec., 2005) 1105--1134,
  [\href{http://xxx.lanl.gov/abs/astro-ph/}{{\tt astro-ph/}}].

\bibitem{Villaescusa_2014a}
F.~{Villaescusa-Navarro}, M.~{Viel}, K.~K. {Datta}, and T.~R. {Choudhury}, {\it
  {Modeling the neutral hydrogen distribution in the post-reionization
  Universe: intensity mapping}},  {\em \jcap} {\bf 9} (Sept., 2014) 50,
  [\href{http://xxx.lanl.gov/abs/1405.6713}{{\tt arXiv:1405.6713}}].

\bibitem{Springel-Hernquist_2003}
V.~{Springel} and L.~{Hernquist}, {\it {Cosmological smoothed particle
  hydrodynamics simulations: a hybrid multiphase model for star formation}},
  {\em \mnras} {\bf 339} (Feb., 2003) 289--311,
  [\href{http://xxx.lanl.gov/abs/astro-ph/0206393}{{\tt astro-ph/0206393}}].

\bibitem{viel13}
M.~{Viel}, G.~D. {Becker}, J.~S. {Bolton}, and M.~G. {Haehnelt}, {\it {Warm
  dark matter as a solution to the small scale crisis: New constraints from
  high redshift Lyman-{$\alpha$} forest data}},  {\em \prd} {\bf 88} (Aug.,
  2013) 043502, [\href{http://xxx.lanl.gov/abs/1306.2314}{{\tt
  arXiv:1306.2314}}].

\bibitem{CAMB}
A.~{Lewis}, A.~{Challinor}, and A.~{Lasenby}, {\it {Efficient Computation of
  Cosmic Microwave Background Anisotropies in Closed Friedmann-Robertson-Walker
  Models}},  {\em \apj} {\bf 538} (Aug., 2000) 473--476,
  [\href{http://xxx.lanl.gov/abs/astro-ph/}{{\tt astro-ph/}}].

\bibitem{Viel_WDM_2012}
M.~{Viel}, K.~{Markovi{\v c}}, M.~{Baldi}, and J.~{Weller}, {\it {The
  non-linear matter power spectrum in warm dark matter cosmologies}},  {\em
  \mnras} {\bf 421} (Mar., 2012) 50--62,
  [\href{http://xxx.lanl.gov/abs/1107.4094}{{\tt arXiv:1107.4094}}].

\bibitem{FoF}
M.~{Davis}, G.~{Efstathiou}, C.~S. {Frenk}, and S.~D.~M. {White}, {\it {The
  evolution of large-scale structure in a universe dominated by cold dark
  matter}},  {\em \apj} {\bf 292} (May, 1985) 371--394.

\bibitem{Subfind}
V.~{Springel}, S.~D.~M. {White}, G.~{Tormen}, and G.~{Kauffmann}, {\it
  {Populating a cluster of galaxies - I. Results at [formmu2]z=0}},  {\em
  \mnras} {\bf 328} (Dec., 2001) 726--750,
  [\href{http://xxx.lanl.gov/abs/astro-ph/}{{\tt astro-ph/}}].

\bibitem{Dolag_2009}
K.~{Dolag}, S.~{Borgani}, G.~{Murante}, and V.~{Springel}, {\it {Substructures
  in hydrodynamical cluster simulations}},  {\em \mnras} {\bf 399} (Oct., 2009)
  497--514, [\href{http://xxx.lanl.gov/abs/0808.3401}{{\tt arXiv:0808.3401}}].

\bibitem{Schneider_2013}
A.~{Schneider}, R.~E. {Smith}, and D.~{Reed}, {\it {Halo mass function and the
  free streaming scale}},  {\em \mnras} {\bf 433} (Aug., 2013) 1573--1587,
  [\href{http://xxx.lanl.gov/abs/1303.0839}{{\tt arXiv:1303.0839}}].

\bibitem{Sheth-Tormen}
R.~K. {Sheth} and G.~{Tormen}, {\it {An excursion set model of hierarchical
  clustering: ellipsoidal collapse and the moving barrier}},  {\em \mnras} {\bf
  329} (Jan., 2002) 61--75, [\href{http://xxx.lanl.gov/abs/astro-ph/}{{\tt
  astro-ph/}}].

\bibitem{Bagla_2010}
J.~S. {Bagla}, N.~{Khandai}, and K.~K. {Datta}, {\it {HI as a probe of the
  large-scale structure in the post-reionization universe}},  {\em \mnras} {\bf
  407} (Sept., 2010) 567--580, [\href{http://xxx.lanl.gov/abs/0908.3796}{{\tt
  arXiv:0908.3796}}].

\bibitem{Dave_2013}
R.~{Dav{\'e}}, N.~{Katz}, B.~D. {Oppenheimer}, J.~A. {Kollmeier}, and D.~H.
  {Weinberg}, {\it {The neutral hydrogen content of galaxies in cosmological
  hydrodynamic simulations}},  {\em \mnras} {\bf 434} (Sept., 2013) 2645--2663,
  [\href{http://xxx.lanl.gov/abs/1302.3631}{{\tt arXiv:1302.3631}}].

\bibitem{Becker_2013}
G.~D. {Becker}, P.~C. {Hewett}, G.~{Worseck}, and J.~X. {Prochaska}, {\it {A
  refined measurement of the mean transmitted flux in the Ly{$\alpha$} forest
  over 2 {\textless} z {\textless} 5 using composite quasar spectra}},  {\em
  \mnras} {\bf 430} (Apr., 2013) 2067--2081,
  [\href{http://xxx.lanl.gov/abs/1208.2584}{{\tt arXiv:1208.2584}}].

\bibitem{Blitz_2006}
L.~{Blitz} and E.~{Rosolowsky}, {\it {The Role of Pressure in GMC Formation II:
  The H$_{2}$-Pressure Relation}},  {\em \apj} {\bf 650} (Oct., 2006) 933--944,
  [\href{http://xxx.lanl.gov/abs/astro-ph/0605035}{{\tt astro-ph/0605035}}].

\bibitem{Krumholz_2011}
M.~R. {Krumholz} and N.~Y. {Gnedin}, {\it {A Comparison of Methods for
  Determining the Molecular Content of Model Galaxies}},  {\em \apj} {\bf 729}
  (Mar., 2011) 36, [\href{http://xxx.lanl.gov/abs/1011.4065}{{\tt
  arXiv:1011.4065}}].

\bibitem{Noterdaeme_2012}
P.~{Noterdaeme}, P.~{Petitjean}, W.~C. {Carithers}, I.~{P{\^a}ris},
  A.~{Font-Ribera}, S.~{Bailey}, E.~{Aubourg}, D.~{Bizyaev}, G.~{Ebelke},
  H.~{Finley}, J.~{Ge}, E.~{Malanushenko}, V.~{Malanushenko},
  J.~{Miralda-Escud{\'e}}, A.~D. {Myers}, D.~{Oravetz}, K.~{Pan}, M.~M.
  {Pieri}, N.~P. {Ross}, D.~P. {Schneider}, A.~{Simmons}, and D.~G. {York},
  {\it {Column density distribution and cosmological mass density of neutral
  gas: Sloan Digital Sky Survey-III Data Release 9}},  {\em \aap} {\bf 547}
  (Nov., 2012) L1, [\href{http://xxx.lanl.gov/abs/1210.1213}{{\tt
  arXiv:1210.1213}}].

\bibitem{Zafar_2013}
T.~{Zafar}, C.~{P{\'e}roux}, A.~{Popping}, B.~{Milliard}, J.-M. {Deharveng},
  and S.~{Frank}, {\it {The ESO UVES advanced data products quasar sample. II.
  Cosmological evolution of the neutral gas mass density}},  {\em \aap} {\bf
  556} (Aug., 2013) A141, [\href{http://xxx.lanl.gov/abs/1307.0602}{{\tt
  arXiv:1307.0602}}].

\bibitem{Noterdaeme_2009}
P.~{Noterdaeme}, P.~{Petitjean}, C.~{Ledoux}, and R.~{Srianand}, {\it
  {Evolution of the cosmological mass density of neutral gas from Sloan Digital
  Sky Survey II - Data Release 7}},  {\em \aap} {\bf 505} (Oct., 2009)
  1087--1098, [\href{http://xxx.lanl.gov/abs/0908.1574}{{\tt
  arXiv:0908.1574}}].

\bibitem{Prochaska_2009}
J.~X. {Prochaska} and A.~M. {Wolfe}, {\it {On the (Non)Evolution of H I Gas in
  Galaxies Over Cosmic Time}},  {\em \apj} {\bf 696} (May, 2009) 1543--1547,
  [\href{http://xxx.lanl.gov/abs/0811.2003}{{\tt arXiv:0811.2003}}].

\bibitem{Villaescusa_2014b}
F.~{Villaescusa-Navarro}, M.~{Viel}, D.~{Alonso}, K.~K. {Datta}, P.~{Bull}, and
  M.~G. {Santos}, {\it {Cross-correlating 21cm intensity maps with Lyman Break
  Galaxies in the post-reionization era}},  {\em ArXiv e-prints} (Oct., 2014)
  [\href{http://xxx.lanl.gov/abs/1410.7393}{{\tt arXiv:1410.7393}}].

\bibitem{Pritchard_2015}
J.~{Pritchard}, K.~{Ichiki}, A.~{Mesinger}, R.~B. {Metcalf}, A.~{Pourtsidou},
  M.~{Santos}, F.~{Abdalla}, T.-C. {Chang}, X.~{Chen}, J.~{Weller}, and
  S.~{Zaroubi}, {\it {Cosmology from the EoR/Cosmic Dawn with the SKA}},  {\em
  ArXiv e-prints} (Jan., 2015) [\href{http://xxx.lanl.gov/abs/1501.0429}{{\tt
  arXiv:1501.0429}}].

\bibitem{Kang_2013}
X.~{Kang}, A.~V. {Macci{\`o}}, and A.~A. {Dutton}, {\it {The Effect of Warm
  Dark Matter on Galaxy Properties: Constraints from the Stellar Mass Function
  and the Tully-Fisher Relation}},  {\em \apj} {\bf 767} (Apr., 2013) 22,
  [\href{http://xxx.lanl.gov/abs/1208.0008}{{\tt arXiv:1208.0008}}].

\end{thebibliography}\endgroup

\end{document}